\documentclass[useAMS,usenatbib]{mn2e}

\usepackage{aas_macros} 
\usepackage{graphicx} 
\usepackage{tabularx} 
\usepackage{multirow} 
\usepackage{rotating}
\usepackage{hhline} 
\usepackage{subfigure} 
\usepackage{amssymb} 
\usepackage{flafter} 
\usepackage{bm} 
\usepackage{mdwlist} 
\usepackage{psfrag}

\newcommand{\eg}{e.g. }
\newcommand{\dif}{\mathrm{d}}
\newcommand{\me}{\mathrm{e}}

\title[SZ observations of galaxy clusters out to the virial
radius]{Sunyaev--Zel'dovich observations of galaxy clusters out to the
  virial radius with the Arcminute Microkelvin Imager \footnote{We
    request that any reference to this paper cites `AMI Consortium:
    Zwart et al.~2009'}}

\author[Zwart et al.]
  {AMI Consortium:
  Jonathan T.~L.~Zwart$^{1,2}$\thanks{Issuing author -- e-mail: jtlz2@astro.columbia.edu. $^*$We request that any reference to this paper cites `AMI Consortium: Zwart et al.~2010'.},
  Farhan~Feroz$^1$,
  \newauthor
  Matthew~L.~Davies$^1$,
  Thomas M. O. Franzen$^1$,
  Keith~J.~B.~Grainge$^{1,3}$,
  \newauthor
  Michael~P.~Hobson$^1$,
  Natasha~Hurley-Walker$^1$,
  R\"{u}diger~Kneissl$^{1,4}$,
  \newauthor
  Anthony~N.~Lasenby$^{1,3}$,
  Malak~Olamaie$^1$,
  Guy~G.~Pooley$^1$,
  \newauthor
  Carmen~Rodr\'{\i}guez-Gonz\'alvez$^1$,
  Richard~D.~E.~Saunders$^{1,3}$,
  \newauthor
  Anna~M.~M.~Scaife$^{1,5}$,
  Paul~F.~Scott$^1$,
  Timothy~W.~Shimwell$^1$,
  \newauthor
  David~J.~Titterington$^1$
  and Elizabeth~M.~Waldram$^1$\\
  $^1$Astrophysics Group, Cavendish Laboratory, J.~J.~Thomson Avenue, Cambridge CB3 0HE\\
  $^2$Columbia Astrophysics Laboratory, Columbia University, 550 West 120th Street, New York, NY 10027, USA\\
  $^3$Kavli Institute for Cosmology Cambridge, Madingley Road, Cambridge CB3 0HA\\
  $^4$Joint ALMA Office, Av El Golf 40, Piso 18, Santiago, Chile\\
  $^5$Dublin Institute for Advanced Studies, 31 Fitzwilliam Place, Dublin 2, Ireland}

\begin{document}

\date{Accepted ---. Received ---; in original form \today.}

\pagerange{\pageref{firstpage}--\pageref{lastpage}}
\pubyear{2010}

\label{firstpage}

\maketitle


\begin{abstract}
  We present observations using the Small Array of the Arcminute
  Microkelvin Imager (AMI; 14--18~GHz) of four Abell and three MACS
  clusters spanning 0.171--0.686 in redshift. We detect
  Sunyaev-Zel'dovich (SZ) signals in five of these without any attempt
  at source subtraction, although strong source contamination is
  present. With radio-source measurements from high-resolution
  observations, and under the assumptions of spherical $\beta$-model,
  isothermality and hydrostatic equilibrium, a Bayesian analysis of
  the data in the visibility plane detects extended SZ decrements in
  all seven clusters over and above receiver noise, radio sources and
  primary CMB imprints. Bayesian evidence ratios range from
  $10^{11}$:1 to $10^{43}$:1 for six of the clusters and 3000:1 for
  one with substantially less data than the others. We present
  posterior probability distributions for, e.g., total mass and gas
  fraction averaged over radii internal to which the mean overdensity
  is 1000, 500 and 200, $r_{200}$ being the virial radius. Reaching
  $r_{200}$ involves some extrapolation for the nearer clusters but
  not for the more-distant ones. We find that our estimates of gas
  fraction are low (compared with most in the literature) and decrease
  with increasing radius. These results appear to be consistent with
  the notion that gas temperature in fact falls with distance (away
  from near the cluster centre) out to the virial radius.
\end{abstract}


\section{Introduction}
\label{sec:intro}

The Sunyaev--Zel'dovich (SZ) effect \citep{sz70,sz72} is the
inverse-Compton scattering of the CMB radiation by hot, ionised gas in
the gravitational potential well of a cluster of galaxies; for reviews
see \cite{birkinshaw99} and \cite{carlstrom02}. The effect is useful
in a number of ways for the study of galaxy clusters; here we are
concerned with two in particular. First, because the SZ effect arises
from a scattering process, a cluster at one redshift will produce the
same observed SZ surface brightness as an identical cluster at any
other redshift, so that the usual sensitivity issue of high-redshift
observing does not arise. Second, since the SZ surface brightness is
proportional to the line-of-sight integral of pressure through the
cluster, the SZ signal is less sensitive to concentration than the
X-ray Bremmsstrahlung signal; one corollary of this is that the ratio
SZ-sensitivity / X-ray-sensitivity increases with distance from the
cluster centre so that with SZ one can probe out to, say, the virial
radius, provided the SZ telescope is sensitive to sufficiently large
angular scales.

SZ decrements are faint, however, and can be contaminated or
obliterated by other sources of radio emission.  A range of new,
sensitive instruments has been brought into use to capitalise on the
science from SZ observations. Among these instruments, which employ
different strategies to maximise sensitivity and minimise confusion,
are ACT \citep{swetz10, menanteau10}, AMI \citep{paper-zero} AMiBA
\citep{ho09,wu09}, APEX \citep{dobbs06}, CARMA
(\texttt{www.mmarray.org}), SPT \citep{spt,andersson10} and SZA
\citep{culverhouse10}. In the case of AMI, two separate interferometer
arrays are used, the Small Array (SA) having short baselines sensitive
to SZ and radio sources, and the Large Array (LA) with baselines
sensitive to the radio sources alone and thus providing source
subtraction for the SA\@. Key parameters of the SA and LA are shown in
Table~\ref{table:ami}.

The SA was built first. Partly to test it while the LA was being
completed, we used the SA to observe Galactic supernova remnants and
likely regions of spinning dust (\citealt{scaife08},
\citeauthor{scaife09b} \citeyear{scaife09a},b, \citeauthor{nhw09a}
\citeyear{nhw09b},b) bright enough not to need source subtraction. But
we also wanted to begin SZ observation, test our algorithms to extract
SZ signals in the presence of radio sources, CMB primary anisotropies
and receiver noise, and begin our SZ science programme. To do this
required the use of long-baseline data from the 15-GHz Ryle Telescope
(RT; see \eg \citealt{grainge93}, \citealt{grainge96},
\citeauthor{grainge02a} \citeyear{grainge02b}a,b,
\citeauthor{cotter02a} \citeyear{cotter02b}a,b, \citealt{grainger02},
\citealt{saunders03}, \citealt{jones05}) taken in the past; this needs
caution because of radio source variability (see \eg
\citealt{bolton06}, \citealt{sadler06} and \citealt{franzen09}), but
our data-analysis algorithm allows for variability and in fact we were
able to use some data from the LA, which, at the time, was only
partially commissioned.  Here we present the first part of this work,
SZ measurements of seven known clusters spanning ranges of redshift
$z$ and of X-ray luminosity $L_X$.

We assume a concordance $\Lambda$CDM cosmology, with
$\Omega_{\mathrm{m}}=0.3$, $\Omega_{\Lambda}=0.7$ and
$H_{0}=70\;\mathrm{km}\;\mathrm{s}^{-1}\;\mathrm{Mpc}^{-1}$. However,
in plots of probability distribution, we explicitly include the
dimensionless Hubble parameter, defined as
$h~\equiv~H_0/\left(100\,\mathrm{kms}^{-1}~\mathrm{Mpc}^{-1}\right)$,
to allow comparison with other work.
All coordinates are J2000 epoch. Our convention for spectral index
$\alpha$ is $S_{\nu}\propto \nu^{-\alpha}$ where $S$ is flux density
and $\nu$ is frequency. We write the radius internal to which the
average density is $a$ times the critical density
$\rho_{\mathrm{crit}}$ at the particular redshift as $r_a$, the total
mass (gas plus dark matter) internal to $r_a$ as $M_a$, and the gas
mass internal to $r_a$ as $M_{\mathrm{gas},a}$.

\begin{table}
\centering
\caption{AMI \citep{paper-zero} technical summary.}\label{table:ami}
\begin{tabular}{{l}{c}{c}}
\hline
                           & SA                   & LA                      \\
\hline
Antenna diameter           & 3.7~m                 & 12.8~m                 \\
Number of antennas         & 10                    & 8                      \\
Baseline lengths (current) & 5--20~m               & 18--110~m              \\
Primary beam (15.7~GHz)    & $20\farcm1$           & $5\farcm5$             \\
Synthesized beam           & $\approx 3\arcmin$   & $\approx 30\arcsec$    \\
Flux sensitivity           & 30~mJy~s$^{1/2}$       & 3~mJy~s$^{1/2}$       \\
Observing frequency        & \multicolumn{2}{c}{13.9--18.2{~GHz}}           \\
Bandwidth                  & \multicolumn{2}{c}{4.3{~GHz}}                  \\
Number of channels         & \multicolumn{2}{c}{6}                          \\
Channel bandwidth          & \multicolumn{2}{c}{0.72{~GHz}}                 \\
\hline
\end{tabular}
\end{table}

\section{Cluster selection and RT observation}
\label{earlies:sample}

We used the NOrthern ROSAT All-Sky Survey (NORAS, \citealt{noras})
catalogue as a source of low-redshift ($z < 0.3$) clusters, and the
MAssive Cluster Survey (MACS, \citealt{MACS}, \citealt{ebeling07},
\citealt{ebeling10}) to give secure, more-distant clusters that
provide some filling-out of the $L_X$--$z$ plane. We restricted
redshifts to $z>0.1$ to avoid resolving out SZ signals, and luminosity
to $L_{\mathrm{X}}>7\times 10^{37}$\,W ($0.1$--$2.4$~keV, rest frame).

We restricted declinations to greater than $20\degr$ since the RT had
only East-West baselines, and further excluded clusters which we knew,
from the NVSS \citep{nvss} or from archival RT data, would be too
contaminated by radio sources. Details of the resulting seven clusters
in this work are given in Table~\ref{tab:clusters}. Source surveying
of the remaining clusters with the compact array of the RT -- note
that this array contained five of the eight antennas of the LA -- was
then carried out as follows.

The RT data were obtained between 2004 and 2006. Each cluster field
was surveyed in two ways: with a wide shallow raster and a deep
central one. The wide shallow raster comprised a hexagonal
close-packed raster of $11\times12$ pointings on a $5\arcmin$ grid,
with a dwell time at each pointing of eight minutes; the aim was to
identify relatively bright radio sources in the direction of an SA
pointing.  The centre of each cluster was followed up with a hexagon
of $7\times 12$-hour RT pointings, on a $5\arcmin$ grid, in order to
detect faint sources near the target cluster.

Data were reduced, and point-source positions and fluxes extracted,
using procedures developed for the 9C survey and outlined in
\citet{waldram03}. The source data are given in Table~\ref{tab:sources}.

\begin{table*}
\centering
\caption{Clusters in this work. Temperatures, redshifts and X-ray luminosities are from $^1$\citet{laroque06}, $^2$ \citet{balestra07} $^3$ \citet{bonamente08}
  $^4$ \citet{ebeling07}, $^5$\citet{noras}, $^6$\citet{struble99}, $^7$ Ebeling (Priv.~Comm.). The map noise
  indicated is for a SA naturally-weighted map with all baselines and no source subtraction. The
  integration times $t_{\mathrm{int}}$ are on-sky times, and do not account for variations in system
  temperature with airmass or poor weather, or for the amount of data flagged due to, for example, shadowing.}
\begin{tabular}{lllcccccc}
\hline
Cluster	&RA (J2000) &Dec (J2000) &$z$ &$T_{\mathrm{e}}$/keV &$L_{X}/10^{37}$W &$t_{\mathrm{int}}$/hours &$\mathrm{rms}$/$\mu$Jy\\
\hline
A611         & 08 00 59.40 & +36 03 01.0 & 0.288 (5) & $6.79^{+0.41}_{-0.38}\,(1)$ & 8.63 (5) & 23.8 & 140 \\
A773         & 09 17 52.97 & +51 43 55.5 & 0.217 (6) & $8.16^{+0.56}_{-0.52}\,(1)$ & 12.11 (5) & 23.8 & 160 \\
A1914        & 14 26 02.15 & +37 50 05.8 & 0.171 (5) & $9.48^{+0.35}_{-0.29}\,(1)$ & 15.91 (5) & 20.9 & 140 \\
A2218        & 16 35 52.80 & +66 12 50.0 & 0.171 (5) & $7.80^{+0.41}_{-0.37}\,(1)$ & 8.16 (5?) & 62.4 & 90 \\
MACSJ0308+26 & 03 08 55.40 & +26 45 39.0 (7) & 0.352 (7) & $11.2^{+0.7}_{-0.7}\,(2)$ & 15.89 (7) & 86.6 & 140 \\
MACSJ0717+37 & 07 17 30.00 & +37 45 00.0 (7) & 0.545 (4) & $11.6^{+0.5}_{-0.5}\,(4)$ & 25.33 (7) & 23.8 & 160 \\
MACSJ0744+39 & 07 44 48.00 & +39 27 00.0 (7) & 0.686 (4) & $8.14^{+0.80}_{-0.72}\,(1)$ & 17.16 (7) & 71.8 & 320 \\
\hline
\end{tabular}
\label{tab:clusters} 
\end{table*}

\begin{table*}
\centering
\caption{
  Contaminating sources.
  W denotes RT wide, shallow raster ($11\times12$
  pointings), while H denotes a RT deep hexagon (7~pointings).
  Fluxes from RT shallow raster observations were boosted by 10 per~cent to account
  for pointing errors \citep{waldram03}.  9C denotes data from 9C
  pointed observations \citep{waldram03}, with the flux error estimated
  at $<5$ per~cent.}
\begin{tabular}{lllllll}
\hline
Cluster	&  &RA (J2000) &Dec (J2000) & Array & Mode & S/mJy \\
\hline
A611         & 1 & 08 00 43.28      & +36 14 00.9      & SA    &   & $5.5   \pm 1.7   $      \\
             & 2 & 08 00 09.91      & +36 04 15.4      & SA    &   & $4.4   \pm 1.3   $      \\
A773         & 1 & 09 18 38.29      & +51 50 25.0      & SA    &   & $4.4   \pm 0.4   $      \\
             & 2 & 09 17 06.13      & +51 44 54.9      & SA    &   & $3.4   \pm 0.3   $      \\
             & 3 & 09 17 57.02      & +51 45 08.0      & LA    &   & $0.12  \pm 0.01 $      \\
             & 4 & 09 18 01.33      & +51 44 13.1      & LA    &   & $0.32  \pm 0.03  $      \\
             & 5 & 09 17 45.31      & +51 43 04.6      & LA    &   & $0.22  \pm 0.02  $      \\
             & 6 & 09 17 55.58      & +51 43 01.1      & LA    &   & $0.19  \pm 0.02 $      \\
             & 7 & 09 17 50.67      & +51 41 06.1      & LA    &   & $0.31  \pm 0.03  $      \\
A1914        & 1 & 14 25 10.21 (SA) & +37 52 35.1 (SA) & SA/LA &   & $4.2   \pm 0.4   $ (LA) \\
             & 2 & 14 27 24.75 (RT) & +37 46 33.8 (RT) & RT/LA &   & $9.7   \pm 1.0   $ (LA) \\
             & 3 & 14 25 48.02      & +37 47 50.3      & LA    &   & $1.0   \pm 0.3   $      \\
             & 4 & 14 25 40.84      & +37 45 50.4      & LA    &   & $3.7   \pm 0.4   $      \\
             & 5 & 14 25 50.53      & +37 45 10.3      & LA    &   & $0.61  \pm 0.18  $      \\
             & 6 & 14 25 58.53      & +37 44 00.1      & LA    &   & $0.60  \pm 0.18  $      \\
             & (7) & 14 25 50.53    & +37 45 10.3      & SA    &   & $4.3   \pm 1.3   $      \\
A2218        & 1 & 16 35 47.24      & +66 14 46.9      & RT    & H & $1.9   \pm 0.6   $      \\
             & 2 & 16 36 15.74      & +66 14 27.0      & RT    & H & $1.9   \pm 0.6   $      \\
             & 3 & 16 35 22.14      & +66 13 20.6      & RT    & W & $5.6   \pm 1.7   $      \\
             & 4 & 16 33 18.18      & +66 00 50.6      & RT    & W & $10    \pm 3     $      \\
             & 5 & 16 35 39.78      & +65 58 12.0      & RT    & W & $11    \pm 3     $      \\
             & 6 & 16 34 46.36      & +65 55 18.6      & RT    & W & $13    \pm 4     $      \\
             & 7 & 16 37 22.56      & +66 21 18.4      & SA(L) &   & $5.2   \pm 1.6   $      \\
MACSJ0308+26 & 1 & 03 09 42.02      & +26 56 30.3      & 9C    & W & $8     \pm 2     $      \\
             & 2 & 03 08 56.52      & +26 44 54.0      & SA(L) &   & $2.4   \pm 0.7   $      \\
             & 3 & 03 09 40.14      & +26 37 23.6      & SA(L) &   & $2.9   \pm 0.9   $      \\
MACSJ0717+37 & 1 & 07 17 36.09      & +37 45 56.3      & RT    & H & $2.1   \pm 0.3   $      \\
             & 2 & 07 17 35.91      & +37 45 11.2      & RT    & H & $1.8   \pm 0.5   $      \\
             & 3 & 07 17 37.14      & +37 44 23.1      & RT    & H & $3.9   \pm 1.2   $      \\
             & 4 & 07 17 41.06      & +37 43 15.2      & RT    & H & $2.5   \pm 0.8   $      \\
             & 5 & 07 18 10.51      & +37 49 14.6      & SA(L) &   & $18    \pm 6     $      \\
             & 6 & 07 16 35.69      & +37 39 14.2      & SA(L) &   & $4.7   \pm 1.4   $      \\
MACSJ0744+39 & 1 & 07 44 32.95      & +39 32 15.0      & RT    & H & $2.8   \pm 0.2   $      \\
             & 2 & 07 44 22.30      & +39 25 46.5      & RT    & H & $1.1   \pm 0.2   $      \\
             & 3 & 07 43 58.76      & +39 15 02.3      & RT    & W & $52.0  \pm 1.7   $      \\
             & 4 & 07 43 45.99      & +39 14 21.5      & RT    & W & $8.3   \pm 1.7   $      \\
\hline
\end{tabular}
\label{tab:sources}
\end{table*}

\section{AMI observation and reduction}

\begin{figure*}
  \centering
  \includegraphics[width=10cm,origin=br,angle=0]{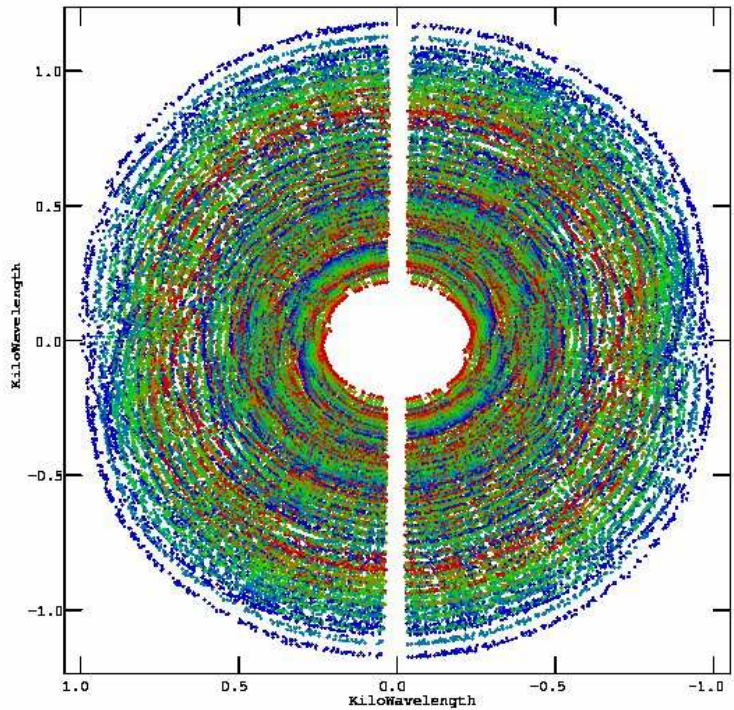}
  \caption{SA \textit{uv}-coverage for A2218; coverages for the other
    clusters are very similar to this. The different colours
    correspond to different frequency channels.\label{fig:uv}}
\end{figure*}

The seven clusters were observed with the SA between 2007 October and
2008 January. Each cluster typically had 25 hours of SA observing on
the sky (though A2218, MACSJ0308+26 and MACSJ0717+27 had some 70
hours). The \textit{uv}-coverage is well-filled (Figure \ref{fig:uv})
all the way down to ${\approx}180{\lambda}$, corresponding to a
maximum angular scale of $\approx 10\arcmin$. This is a significantly
greater angular scale than is achievable with OVRO/BIMA, the RT, or
the SZA.

Calibration and reduction procedures were as follows. One of our two
absolute flux calibrators, 3C286 and 3C48, was observed immediately
before or after each cluster observation. The absolute flux
calibration is accurate to 5 per~cent (see \citealt{nhw09a}). Each
cluster observation was reduced separately using our in-house software
\textsc{reduce}. An automatic reduction pipeline is in place, but all
the data were examined by eye for problems. Data were flagged for
shadowing, slow fringe rates, path-compensator delay errors and
pointing errors. The data were flux-calibrated, Fourier transformed
and fringe-rotated to the pointing centre. Further amplitude cuts were
made in order to remove interference spikes and discrepant
baselines. The amplitudes of the visibilities were corrected for
variations in the system temperature with airmass, cloud and weather,
and the data weights converted into Jy$^{-2}$. Secondary (interleaved)
calibration was applied, by observing a point-source calibrator every
hour, to correct for system phase drifts. The data were smoothed from
one-second to 10-second samples, and calibrated \textsc{uvfits} were
outputted and co-added using \textsc{pyfits}. Typically 20--30 per
cent of the data were discarded due to bad weather, telescope downtime
and other flagging. The data were mapped in \textsc{aips} and also
directly analysed in the visibility plane.

In some cases, as indicated in Table~\ref{tab:sources}, it was
possible to use some of the then partially commissioned LA for source
subtraction, assisting with any effects of the time gap between RT and
SA observations (LA calibration and reduction are very similar to that
of the SA, described above). Similarly, for some sources of high flux
density away from the cluster, the long baselines of the SA provided
useful measurements.

\subsection{Maps}

\begin{figure*}
  \centering
  \mbox{\subfigure[A611. The 1-$\sigma$ map noise is $139\,\mu$Jy. Contour levels start at $\pm 280\,\mu$Jy and increase at each level by a factor of $\sqrt{2}$.]{\label{fig:a611}\includegraphics[width=7cm,origin=br,angle=0]{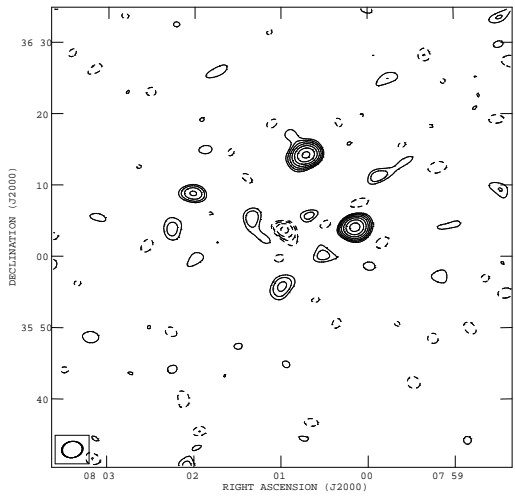}}}\qquad
\mbox{\subfigure[A1914. The 1-$\sigma$ map noise is $144\,\mu$Jy. Contour levels start at $\pm 290\,\mu$Jy and increase at each level by a factor of $\sqrt{2}$.]{\label{fig:a1914}\includegraphics[width=7cm,origin=br,angle=0]{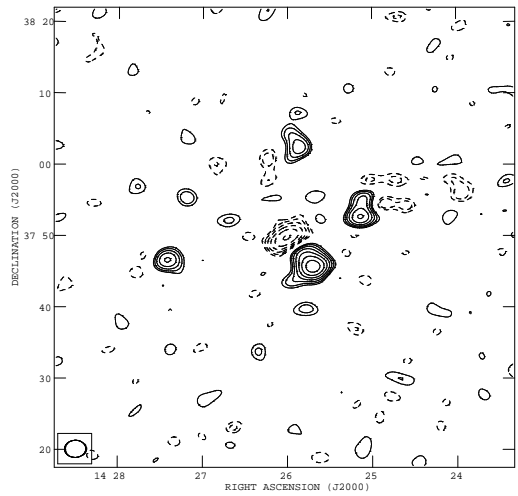}}}\\
  \mbox{\subfigure[A773. The 1-$\sigma$ map noise is $157\,\mu$Jy. Contour levels start at $\pm 310\,\mu$Jy and increase at each level by a factor of $\sqrt{2}$.]{\label{fig:a773}\includegraphics[width=7cm,origin=br,angle=0]{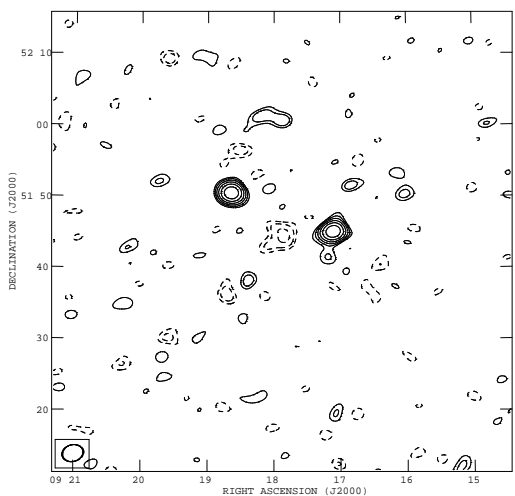}}}\qquad
\mbox{\subfigure[A2218. The 1-$\sigma$ map noise is $88\,\mu$Jy. Contour levels start at $\pm 180\,\mu$Jy and increase at each level by a factor of $\sqrt{2}$.]{\label{fig:a2218}\includegraphics[width=7cm,origin=br,angle=0]{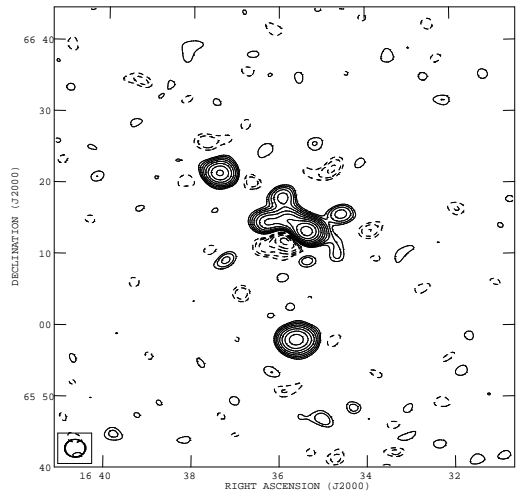}}}\\
\caption{SA naturally-weighted maps of the Abell clusters. No source
  subtraction has been done. The synthesized beam is
  indicated in the lower left corner of each image.\label{fig:abell}}
\end{figure*}

We used standard \textsc{aips} tasks to produce naturally weighted SA
maps with all baselines, no taper and no source subtraction. These
images, after \textsc{clean}ing, are shown in
Figure~\ref{fig:abell}. The maps have differing noises due largely to
differing integration times.  Sources are evident in all the maps. In
five of the maps, an extended SZ decrement is visible, despite major
source contamination at the X-ray centres in the cases of A2218 and
MACSJ0308+26. In MACSJ0717+37, there seems to be some negative signal
but the source contamination at the map centre is severe
\citep{edge03,ebeling04}. In MACSJ0744+39, the contamination is less
but there is still only a weak decrement -- but we note that the
thermal noise is at least twice that of every other map.

Subsequent analysis was carried out in the visibility plane, taking
into account radio sources, receiver noise and primary CMB
contamination, as we describe in the next section.

\section{Resume of analysis}

\subsection{Bayesian analysis}

Bayesian analysis of interferometer observations of clusters in SZ has
been discussed by us previously in e.g. \citet{HM02},
\citet{marshall03} and \citet{mcadam}. The advantages of this approach
are as follows.

\begin{itemize}

\item One infers the quantity that one actually wants, the probability
  distribution of the values of parameters $\mathbf{\Theta}$, given the
  data $\mathrm{\mathbf{D}}$ and some model, or hypothesis, $H$, via
  Bayes' theorem:

\begin{equation} 
\mathrm{Pr}\left(\mathbf{\Theta}|\mathrm{\mathbf{D}},H\right)
= \frac{
\mathrm{Pr}\left(\mathrm{\mathbf{D}}|\mathbf{\Theta},H\right)
\mathrm{Pr}\left(\mathbf{\Theta}| H\right)}
{\mathrm{Pr}\left(\mathrm{\mathbf{D}}| H\right)}.
\end{equation} 

\item The likelihood
  $\mathrm{Pr}\left(\mathrm{\mathbf{D}}|\mathbf{\Theta},H\right)$ is
  the probability of the data given parameter values and a model, and
  encodes the constraints imposed by the observations. It includes
  information about noise arising from the receivers, primary CMB and
  unsubtracted radio sources lying below the detection level of the
  source-subtraction procedure.

\item The prior $\mathrm{Pr}\left(\mathbf{\Theta}| H\right)$ allows
  one to incorporate prior knowledge of the parameter values and, for
  example, allows one to deal fully and objectively with the
  contaminants such as sources (which may be variable).

\item The evidence $\mathrm{Pr}\left(\mathrm{\mathbf{D}}| H\right)$ is
  obtained by integrating
  $\mathrm{Pr}\left(\mathrm{\mathbf{D}}|\mathbf{\Theta},H\right)
  \mathrm{Pr}\left(\mathbf{\Theta}| H\right)$ over all
  $\mathbf{\Theta}$, allowing normalization of the posterior
  $\mathrm{Pr}\left(\mathbf{\Theta}|\mathrm{\mathbf{D}},H\right)$. One
  can select different models by comparing their evidences, the
  process automatically incorporating Occam's razor.

\item However, performing these integrations, and sampling the
  parameter space, is non-trivial and can be slow. The use of the
  `nested sampler' algorithm \textsc{MultiNest} both speeds up the
  sampling process significantly and, more importantly, allows one to
  sample from probability distributions with multiple peaks and/or
  large curving degeneracies \citep{feroz08}.

\item Throughout the whole analysis, probability distributions -- with
  their asymmetries, skirts, multiple peaks and whatever else -- are
  used and combined correctly, rather than discarding information
  (and, in general, introducing bias) by representing distributions by
  a mean value and an uncertainty expressed only in terms of a
  covariance matrix.

\end{itemize}

\subsection{Physical Model and Assumptions}

We restrict ourselves to the simplest model, by assuming a spherical
$\beta$-model for isothermal (see section \ref{model:priors}), ideal
cluster gas in hydrostatic equilibrium.  Following
e.g. \cite{grego01}, the equation of hydrostatic equilibrium for a
spherical shell of gas of density $\rho$ at pressure $p$, a radius $r$
from the cluster centre is

\begin{equation}
\label{eqn:hydro2}
\frac
{\dif p \left(r\right)}
{\dif r}
=-\frac{GM_{r}\rho\left(r\right)}{r^2},
\end{equation}

\noindent where $M_{r}\equiv M\left(<r\right)$ is the total mass (gas
plus dark matter) internal to radius $r$ and the gas' density
distribution $\rho\left(r\right)$ is

\begin{equation}
\label{eqn:beta}
\rho\left(r \right)=
\frac{\rho\left(r=0\right)}{\left[1+\left(r/r_c\right)^2 \right]^{3\beta/2}}.
\end{equation}

\noindent The density profile has a flat top at low $r/r_c$ (with
$r_c$ the core radius), then turns over, and at large $r/r_c$ has a
logarithmic slope of $-3\beta$. The profile may be integrated to find
the gas mass $M_{\mathrm{gas}}$ within $r$.

One also requires the equation of state of the gas,
i.e.~$p\left(\rho\right)$. For ideal gas, $p =
\frac{\rho}{\mu}k_{\mathrm{B}}T$, with $\mu$ the effective mass of
protons per gas particle (we take $\mu=0.6m_p$), equation
(\ref{eqn:hydro2}) becomes

\begin{equation}
\label{eqn:hydro3}
\frac{\dif }{\dif r}
\left(
\frac{\rho k_{\mathrm{B}}T}{\mu}
\right)
=-\frac{GM_{r}\rho}{r^2},
\end{equation}

\noindent and one obtains

\begin{equation}
\label{eqn:hydro32}
M_{r}
=
-\frac{k_{\mathrm{B}}T}{\mu G}
\frac{r^2}{\rho}
\frac{\dif \rho}{\dif {r}}
=
\frac{3\beta r^3}{r_c^2 + r^2}
\frac{k_{\mathrm{B}}T}{\mu G}.
\end{equation}  

\subsection{Priors used here}
\label{model:priors}

The forms of the priors we have assumed for cluster and source
parameters are given in Table~\ref{tab:priors}. Positions
$\mathbf{x}_c$, redshifts $z$ and gas temperatures $T_{\mathrm{e}}$
for individual clusters are quoted in Table \ref{tab:clusters}. For
the sources, positions $\mathbf{x}_i$ and fluxes $S_i$ are in Table
\ref{tab:sources}, and $\alpha_i$ is the 15--22~GHz probability kernel
for source spectral index.  Note that for radio sources, we use
$\delta$-functions on source positions since the position error of a
source is much smaller than an SA synthesized beam, while for source
fluxes, we use a Gaussian centred on the flux density from
high-resolution observations with a 1-$\sigma$ width of $\pm$
30~per~cent to allow for variability, but for A773 we later tighten
the prior on source flux (see \citealt{mcadam} for details). We next
comment on our use of a single temperature for each cluster.

\begin{table*}
\centering
\caption{Fitted parameter names and priors for the cluster
  analysis. The 15--22~GHz probability kernel for source spectra is $\alpha_i$. 
  \label{tab:priors}}
\begin{tabular}{ll}
\hline
Cluster: & \\
\hline
$\mathbf{x}_c$  & Gaussian, $\sigma=1.0\arcmin$ \\
$z$           & $\delta$-function \\
$r_{\mathrm{c}}$    & Uniform, 10--1000~kpc $h^{-1}$ \\
$\beta$         & Uniform, 0.3--1.5 \\
$T_{\mathrm{e}}$    & Gaussian, value from literature $\pm15\%$ \\
$M_{\mathrm{gas},200}$  & Uniform in log-space, $(0.01$--$5.00)\times10^{14} M_{\odot}\,h^{-2}$\\
\hline
Radio sources: & \\
\hline
$\mathbf{x}_i$  & $\delta$-function \\
$S_i$           & Gaussian, $\pm$30 per~cent \\
$\alpha_i$      & Smoothed version of that in \cite{waldram07}\\
\hline
\end{tabular}
\end{table*}

Most SZ work so far has concentrated on the inner parts of clusters,
but as one moves to radii larger than, say, $r_{2500}$ the
observational position on $T_{\mathrm{e}}(r)$ seems to be unclear. The
following examples from the literature attempt to measure
$T_{\mathrm{e}}(r)$ out to about half the classical virial radius,
i.e.~half of $r_{180}$ \citep{peebles93}, in samples of clusters. In
30 clusters observed with ASCA, \citet{Mark98} find that on average
$T_{\mathrm{e}}$ drops to about 0.6 of its central value by
0.5$r_{180}$.  Using ROSAT observations of 26 clusters,
\citet{irwin99} rule out a temperature drop of 20 per~cent at 10~keV
within 0.35$r_{180}$ at 99 per~cent confidence.  With
\textit{BeppoSAX} observations of 21 clusters, \citet{degrandi02} find
that on average $T_{\mathrm{e}}$ falls to about 0.7 of its central
value by 0.5$r_{180}$. With \textit{Chandra} obervations of 13 relaxed
clusters, \citet{vik05} find that on average $T_{\mathrm{e}}$ falls by
about 40 per~cent between 0.15$r_{180}$ and 0.5$r_{180}$ but with
near-flat exceptions. In \textit{XMM-Newton} observations of 48
clusters, \citet{lecmol08} find that most have $T_{\mathrm{e}}$
falling by 20--40 per~cent from 0.15$r_{180}$ to 0.4$r_{180}$ but that
a minority are flat. Using \textit{XMM-Newton} data on 37 clusters,
\citet{zhang08} find that $T_{\mathrm{e}}(r)$ is broadly flat between
0.02$r_{500}$ and 1$r_{500}$.

We have tried to find measurements in the literature of
$T_{\mathrm{e}}(r)$ out to large $r$ for our seven clusters, with the
following results. Using \textit{Chandra} data on A611, \citet{don10}
find that $T_{\mathrm{e}}$ peaks at 200~kpc and falls to 80 per~cent
of the peak at 600~kpc. We could not find a radial profile for A773,
but \citet{gov04} show a temperature map from \textit{Chandra} out to
400~kpc radius; assessing this purely by eye, we estimate that the
mean $T_{\mathrm{e}}$ is about 8~keV with hotter and colder patches
but no clear radial trend. For A1914, \citet{zhang08} find from
\textit{XMM-Newton} data that $T_{\mathrm{e}}(r)$ is flat from 150 to
900~kpc, while on the other hand \citet{mrocz09} find from
\textit{Chandra} data that $T_{\mathrm{e}}(r)$ falls from 9~keV at
0.2~Mpc to 6.6~keV at 1.2~Mpc. For A2218, \citet{pratt05} find from
\textit{XMM-Newton} data that $T_{\mathrm{e}}(r)$ falls from 8~keV
near the centre to 6.6~keV at 700~kpc. Unsurprisingly, we have been
unable to find $T_{\mathrm{e}}(r)$ estimates for our MACS clusters,
which are distant.

X-ray analysis at large $r$ is of course hampered by uncertainty in
the background. The satellite \textit{Suzaku} has a low orbit which
results in some particle screening by the Earth's magnetic field and
thus a low background. \citet{george09} find that in cluster
PKS0745-191, $T_{\mathrm{e}}(r)$ falls by roughly 70 per~cent from
0.3$r_{200}$ to $r_{200}$ with no extrapolation of the data in $r$ and
indeed going beyond $r_{200}$, and \citet{bautz09} and \citet{hosh10}
find somewhat similar behaviour in respectively A1795 and A1413. As
far as we know, these are as yet the only relevant X-ray observations
that extend to very large $r$.

In view of the foregoing, we chose to assume isothermality (at the
temperatures given in Table \ref{tab:clusters}), and to examine the
consequences in this case.

\begin{figure*}
  \centering
  \mbox{\subfigure[MACSJ0308+26. The 1-$\sigma$ map noise is $141\,\mu$Jy. Contour levels start at $\pm 280\,\mu$Jy and increase at each level by a factor of $\sqrt{2}$.]{\label{fig:macs0308+26}\includegraphics[width=7cm,origin=br,angle=0]{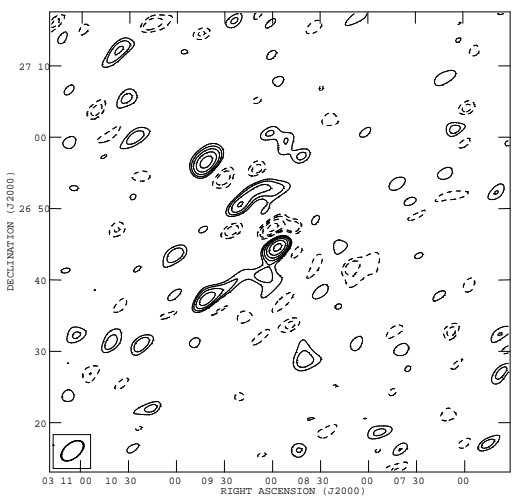}}}\qquad
  \mbox{\subfigure[MACSJ0717+37. The 1-$\sigma$ map noise is $161\,\mu$Jy. Contour levels start at $\pm 320\,\mu$Jy and increase at each level by a factor of $\sqrt{2}$.]{\label{fig:macs0717+37}\includegraphics[width=7cm,origin=br,angle=0]{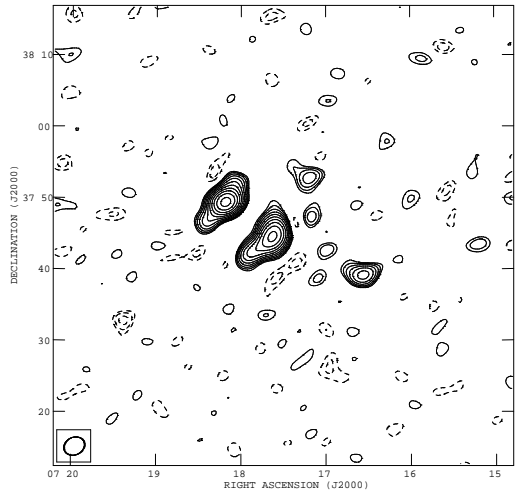}}}\qquad
\mbox{\subfigure[MACSJ0744+39. The 1-$\sigma$ map noise is $317\,\mu$Jy. Contour levels start at $\pm 630\,\mu$Jy and increase at each level by a factor of $\sqrt{2}$.]{\label{fig:macs0744+39}\includegraphics[width=7cm,origin=br,angle=0]{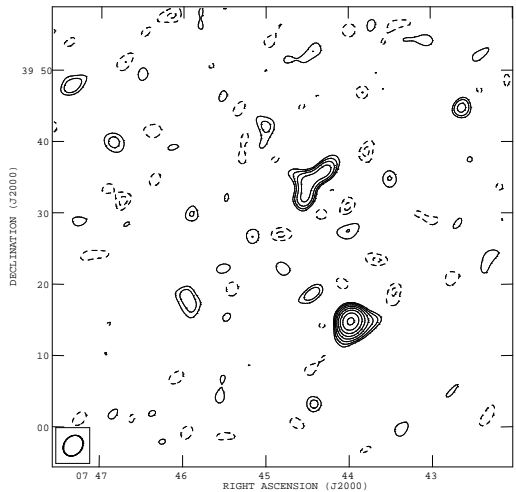}}}\\
\caption{SA naturally-weighted maps of the MACS clusters. No source
  subtraction was undertaken for these images. The synthesized beam is
  indicated in the lower left corner of each image.\label{fig:macs}}
\end{figure*}

\section{Evidences}
\label{results:evidences}

We consider two basic models, as follows.  The first model consists of
hypothesis $H_{1}$ that the data support thermal and CMB noise plus a
number of contaminating radio sources, together with priors on source
parameters.  The second model consists of hypothesis $H_{2}$ that the
data support the two noise contributions plus the contaminating
sources and also a cluster in the SZ with a $\beta$-profile, plus
priors on the fitted parameters. We have carried out the analysis in
two stages: first, determining the best modelling of the source
contributions in each cluster field; and second determining in each
field the extent, if any, to which $H_{2}$ is supported over $H_{1}$.

\subsection{Source model selection}

Inside each of $H_{1}$ and $H_{2}$, we can consider different models
for the field of contaminating sources. We now discuss the use of the
Bayesian evidence for model selection in the two cases (A773 and
A1914) for which source observations had suggested a possible choice
of source model.

\subsubsection{A773}
\label{results:evidences:a773}

The models for A773 all include seven point sources: none was detected
with the RT, two were found in the SA data and five were found with
subsequent LA observations (see Table \ref{tab:sources}). We compared
two models, in which the flux uncertainties were $ \pm 30$ per cent,
to allow for variability, and another in which the flux uncertainties
were reduced to $\pm 10$ per~cent. We carried out a Bayesian analysis
run for the first model and another for the second. The difference in
the $\log_{e}$-evidence was $1.20 \pm 0.11$, marginally favouring the
10-per~cent model; that is, the odds in favour of the 10-per~cent
model over the 30-per~cent model are $3.3 \pm 1.1$ to 1. There is thus
little to choose between the models. For A773 we have used the
10-per~cent model but kept the 30-per~cent model for the other
clusters.

\subsubsection{A1914}
\label{results:evidences:a1914}

For A1914, we consider three source models, all of which have one
source from the SA long baselines and four sources detected with the
LA. In one of the models (A) we include an RT-detected source; in a
second (B), the flux for that source is taken from the LA data (which
were taken much closer in time to the SA observations), and the errors
are tightened; in the third model (C), a further source (source 7)
that is possibly detected by the SA is also included.  The relative
$\log_{e}$-evidences for each model with respect to model C and given
$H_{2}$ are shown in Table \ref{table:a1914}.

\begin{table}
\centering
\caption{Relative evidences 
   for different source models for A1914.\label{table:a1914}}
\begin{tabular}{rrl}
\hline
Model & Sources  & Relative $\log_{\mathrm{e}}$-evidence \\
\hline
A & 6 &  $5.56 \pm 0.19$ \\
B & 6 & $10.05 \pm 0.17$ \\
C & 7 &  0.0 \\
\hline
\end{tabular}
\end{table}

Model C, which includes the source candidate possibly detected by the
SA, is overwhelmingly disfavoured relative to the two models (A and B)
that have only six sources, and we discard model C.

Of the two models with six sources, model B, in which the point-source
flux errors are tightened, is favoured (relative to model A) by an
odds ratio of $\me^{4.49\pm 0.16}$. Consequently we select model B as
the preferred model for parameter estimation. Once again we see that
the Bayesian evidence is a useful and straightforward tool for model
selection in cases where we want to test for source detection and
errors on prior fluxes.

\subsection{Cluster Detections}
\label{results:evidences:all}

For each cluster, the $\log_{e}$-evidence difference $\mathsf{\Delta
  Z}$ for $H_{2}$ over $H_{1}$, that is, the $\log_{e}$-evidence for
an SZ signal over and above (thermal noise plus CMB primary
anisotropies plus the sources) for each cluster model are shown in
Table \ref{tab:evs}. Thus the evidence ratios, given by $E = \exp
{\mathsf{\Delta Z}}$, are huge (ranging from $10^{11}$ to $10^{43}$)
except for MACSJ0744+39. For this cluster, $E$ is about 3000,
i.e. there is a 1 in 3000 chance that the SZ detection is spurious;
note that this is the cluster for which the thermal noise is at least
twice that of any of the others. Of course, we know from optical
and/or X-ray that a cluster is present in each case. Thus the high
$E$-values indicate the power of the observing plus analysis
methodology for detecting SZ even in the presence of serious source
confusion. The methodology works even with substantial uncertainty on
the source fluxes but requires that the existences of the sources, in
approximately the right positions, are correctly determined.

\begin{table}
\centering
\caption{For each cluster, the $\log_{e}$-evidence $\mathsf{\Delta Z}$ for
  an SZ signal in addition to (thermal noise plus CMB primary anisotropies plus 
  the $n$ sources).\label{tab:evs}}
\begin{tabular}{lrrrr}
\hline
  \multicolumn{1}{c}{Cluster} &
 \multicolumn{1}{c}{$n$} &
 \multicolumn{1}{c}{$\Delta Z$} \\
\hline
  A611         & 2 & $27.27 \pm 0.12$ \\ 
  A773         & 7 & $27.13 \pm 0.09$ \\ 
  A1914        & 6 & $64.84 \pm 0.11$ \\ 
  A2218        & 7 & $92.26 \pm 0.23$ \\ 
  MACSJ0308+26 & 3 & $47.59 \pm 0.13$ \\ 
  MACSJ0717+37 & 6 & $33.90 \pm 0.19$ \\ 
  MACSJ0744+39 & 4 & $7.88  \pm 0.16$ \\ 
\hline
\end{tabular}
\end{table}

\section{Parameter Estimates and Discussion}
\label{results:params}

\begin{figure*}
\centering
\subfigure[For fitted parameters, posteriors marginalized into two dimensions, and into one dimension along the diagonal.]{\includegraphics[width=11cm,origin=br,angle=0]{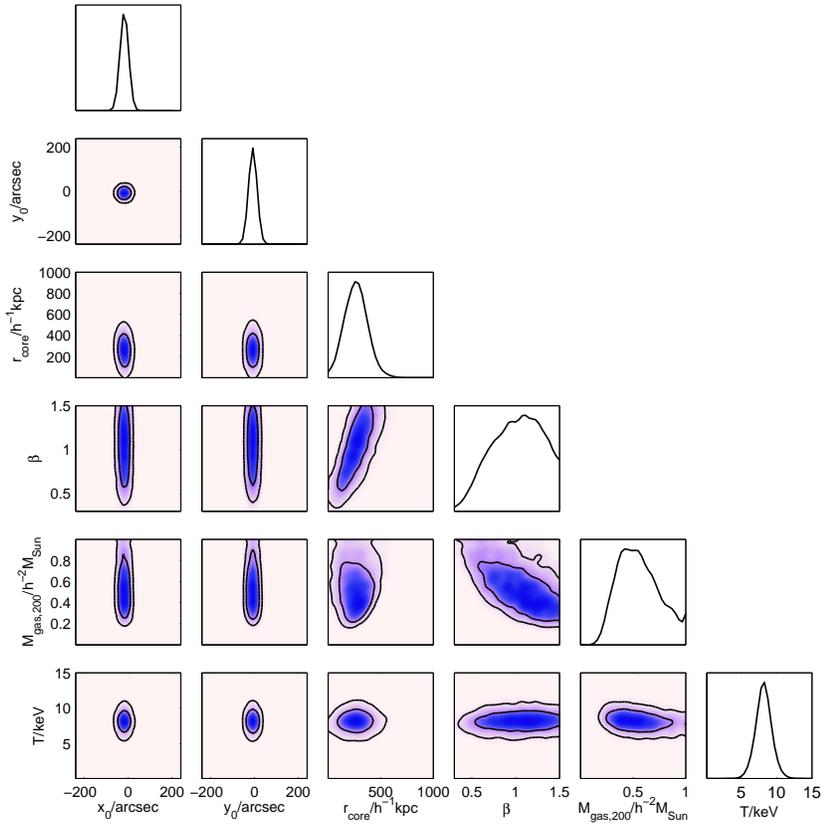}}
\subfigure[For derived parameters, posteriors marginalized into one dimension.]{\includegraphics[trim = 0mm 0mm 0mm 45mm,clip,width=11cm,origin=br,angle=0]{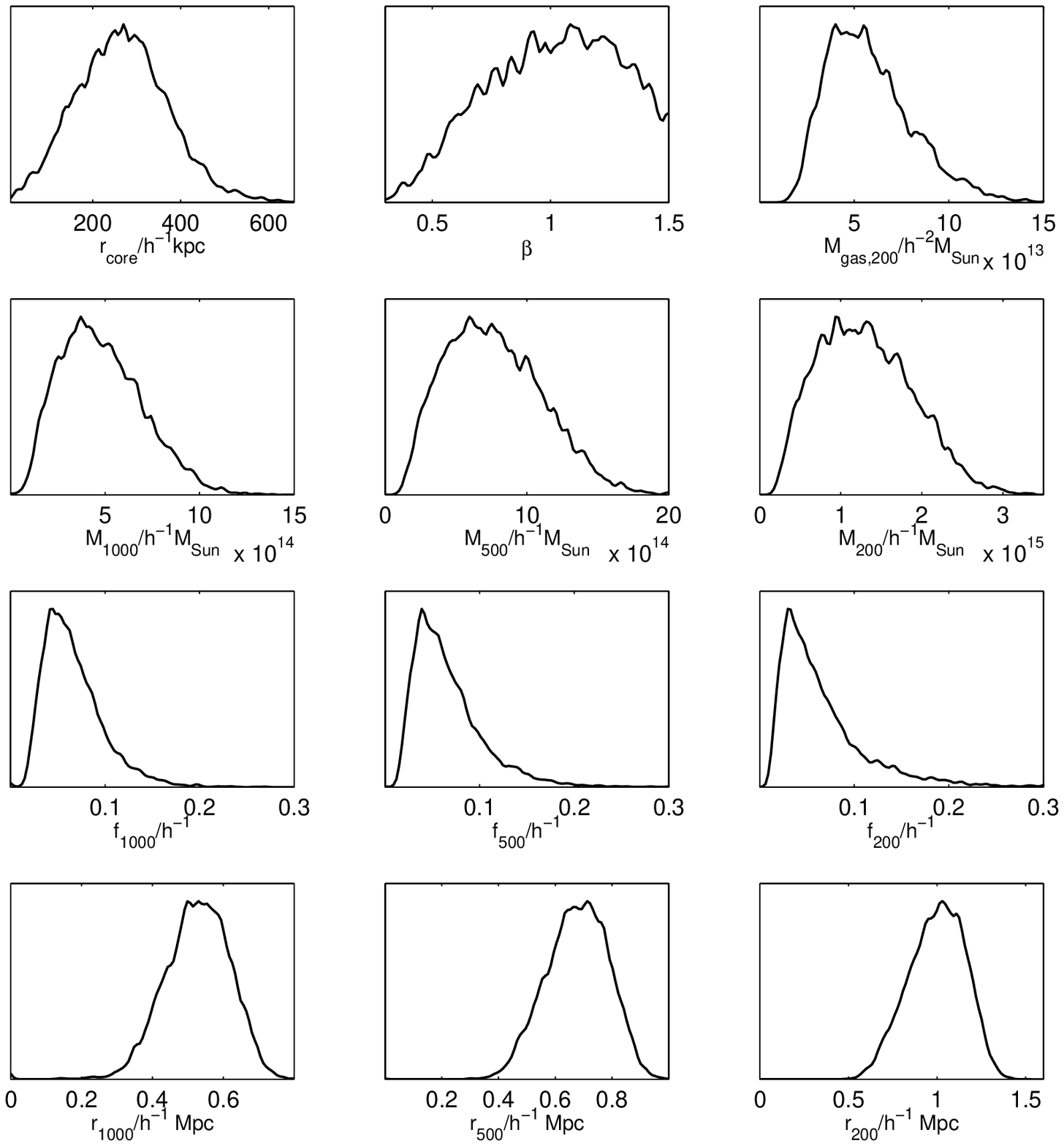}}
\caption{A773 posterior probability distribution.\label{fig:post:a773}}
\end{figure*}

\begin{figure*}
\centering
\subfigure[For fitted parameters, posteriors marginalized into two dimensions, and into one dimension along the diagonal.]{\includegraphics[width=11cm,origin=br,angle=0]{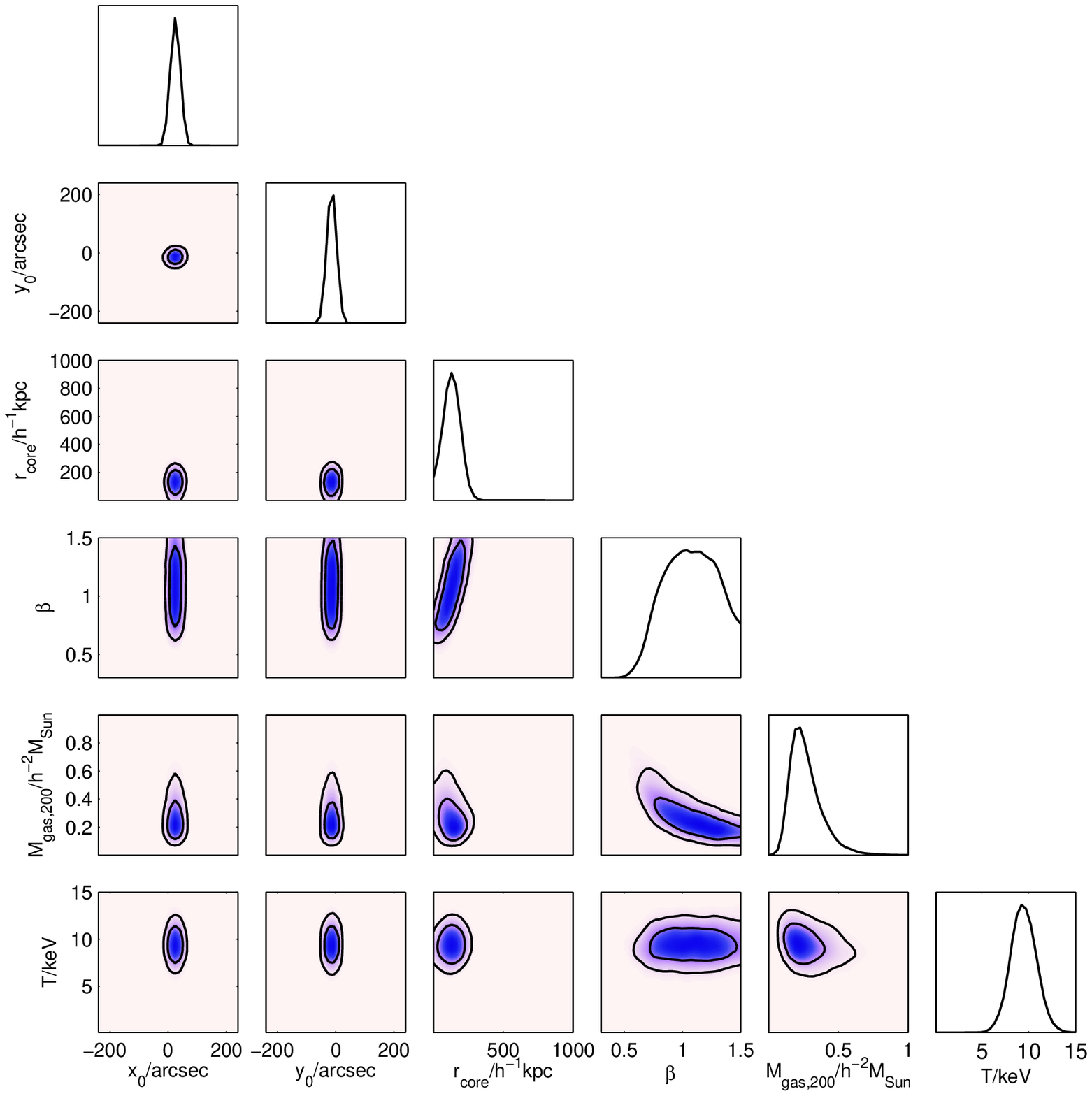}}
\subfigure[For derived parameters, posteriors marginalized into one dimension.]{\includegraphics[trim = 0mm 0mm 0mm 45mm,clip,width=11cm,origin=br,angle=0]{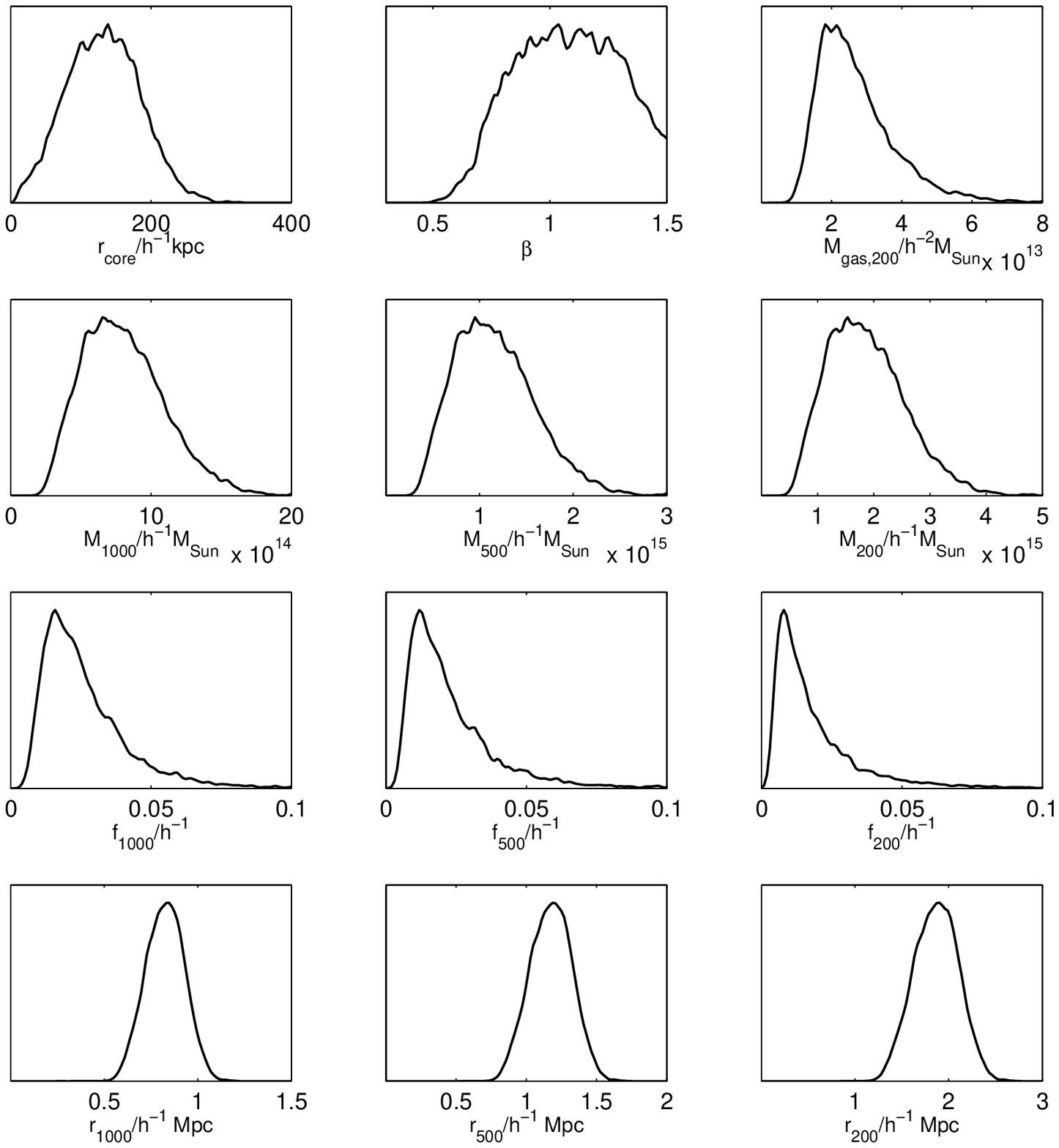}}
\caption{A1914 posterior probability distribution.\label{fig:post:a1914}}
\end{figure*}

\begin{figure*}
\centering
\subfigure[For fitted parameters, posteriors marginalized into two dimensions, and into one dimension along the diagonal.]{\includegraphics[width=11cm,origin=br,angle=0]{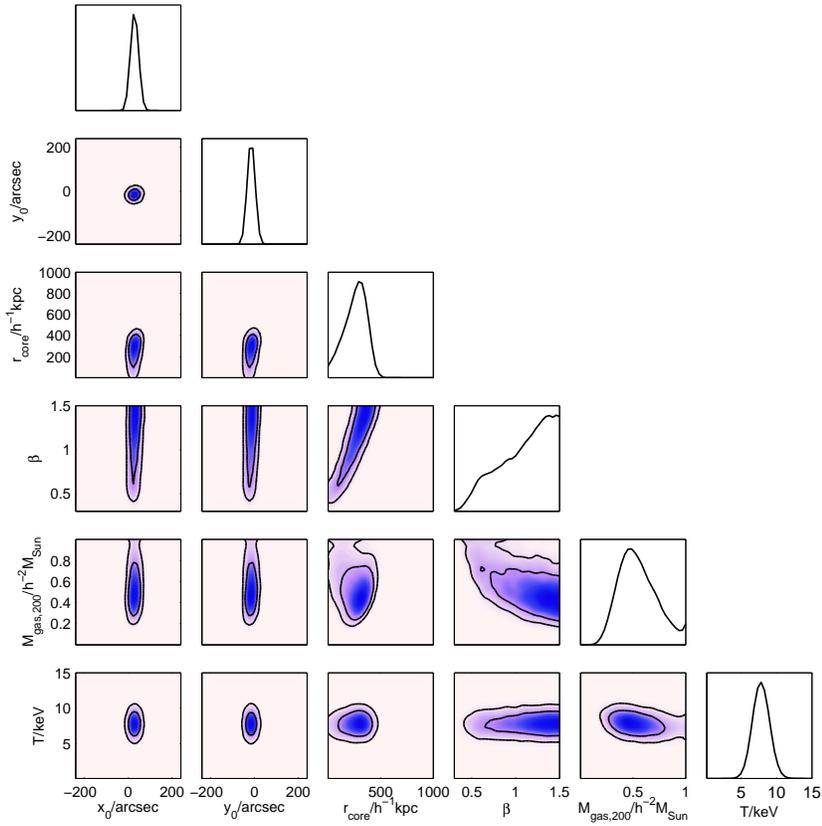}}
\subfigure[For derived parameters, posteriors marginalized into one dimension.]{\includegraphics[trim = 0mm 0mm 0mm 45mm,clip,width=11cm,origin=br,angle=0]{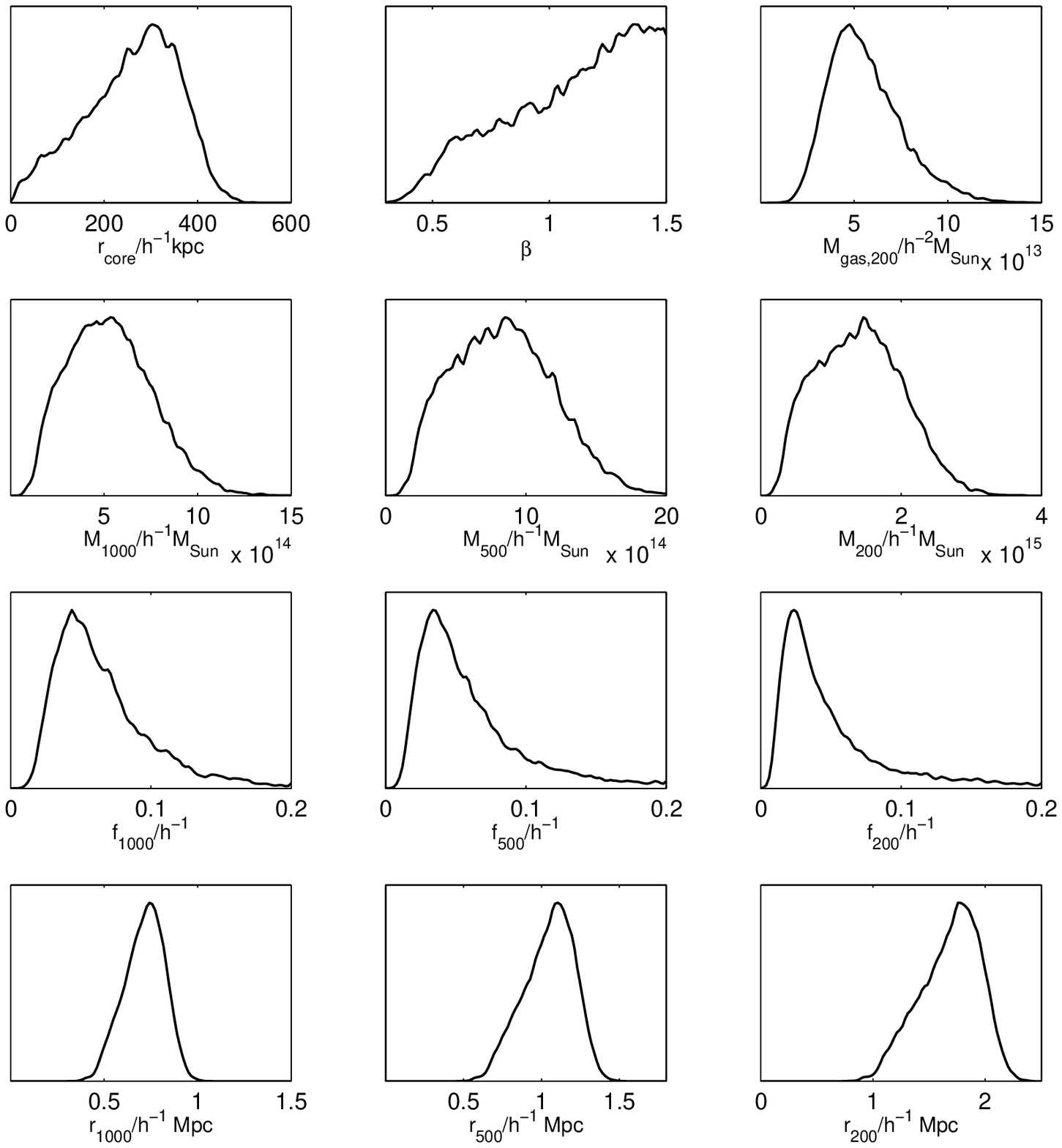}}
\caption{A2218 posterior probability distribution.\label{fig:post:a2218}}
\end{figure*}

\begin{figure*}
\centering
\subfigure[For fitted parameters, posteriors marginalized into two dimensions, and into one dimension along the diagonal.]{\includegraphics[width=11cm,origin=br,angle=0]{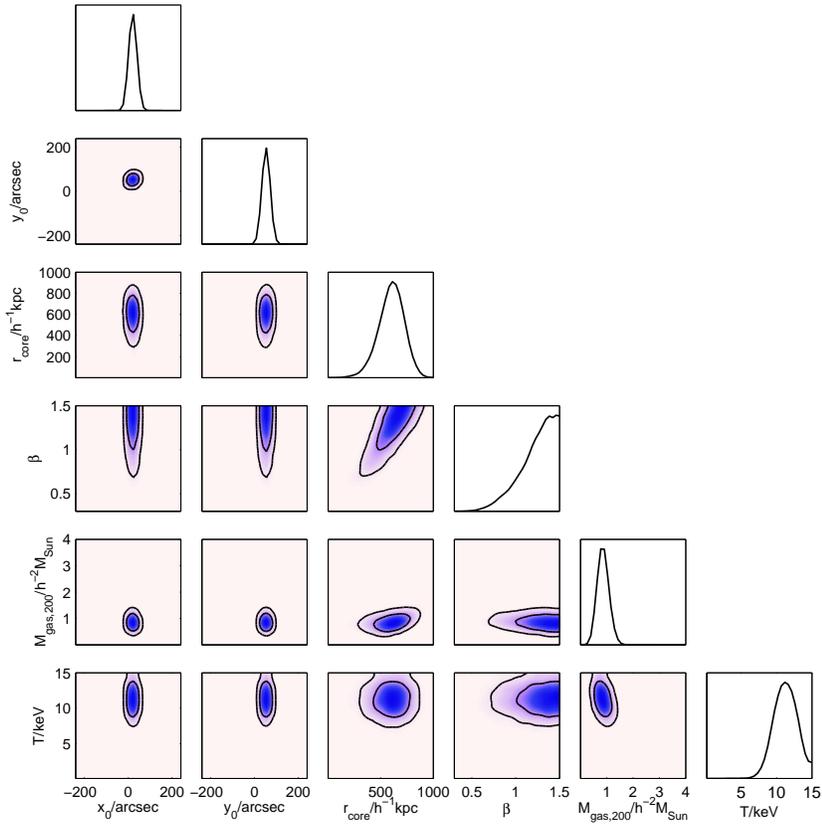}}
\subfigure[For derived parameters, posteriors marginalized into one dimension.]{\includegraphics[trim = 0mm 0mm 0mm 45mm,clip,width=11cm,origin=br,angle=0]{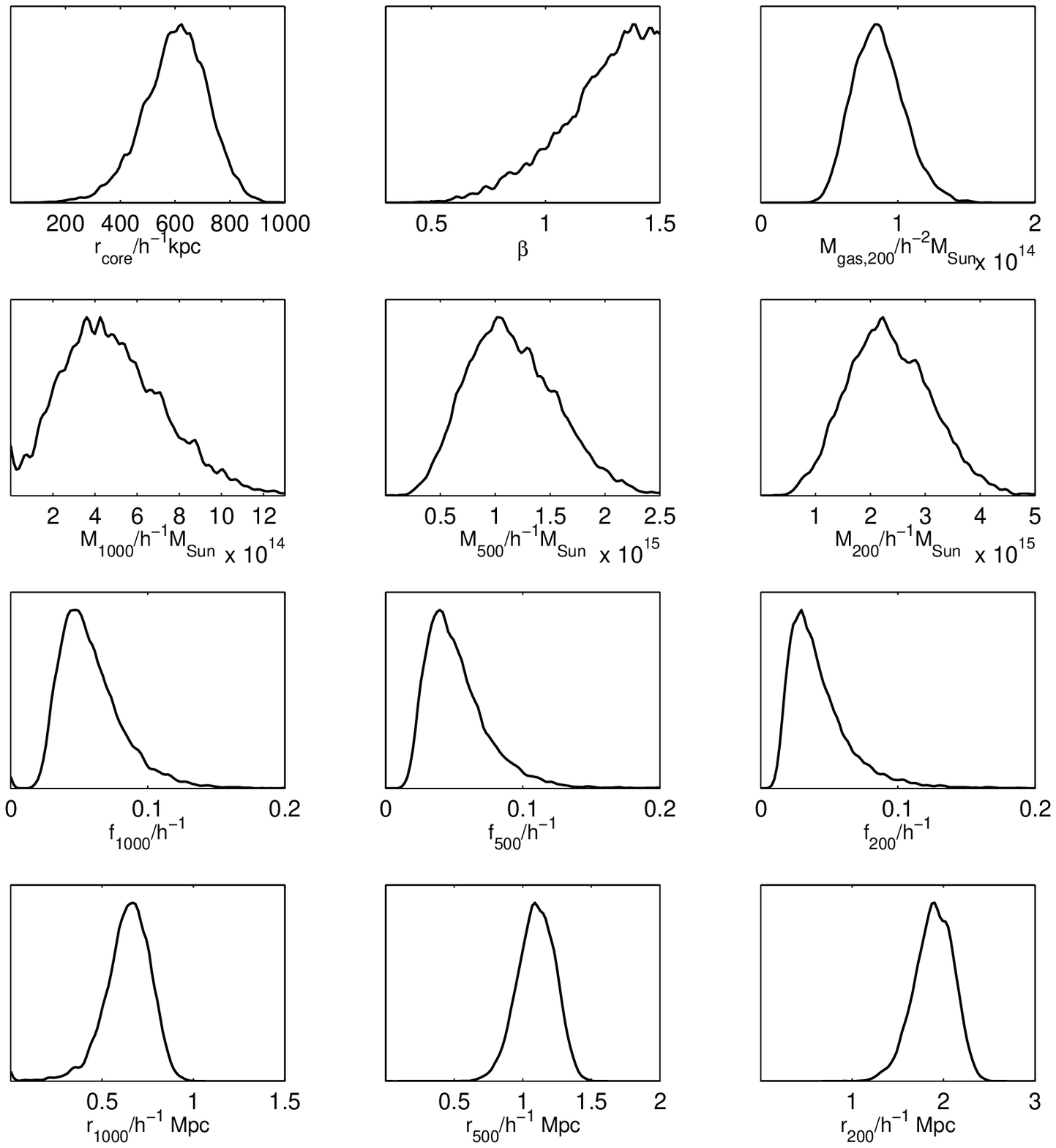}}
\caption{MACSJ0308+26 posterior probability distribution.\label{fig:post:0308}}
\end{figure*}

\begin{figure*}
\centering
\subfigure[For fitted parameters, posteriors marginalized into two dimensions, and into one dimension along the diagonal.]{\includegraphics[width=11cm,origin=br,angle=0]{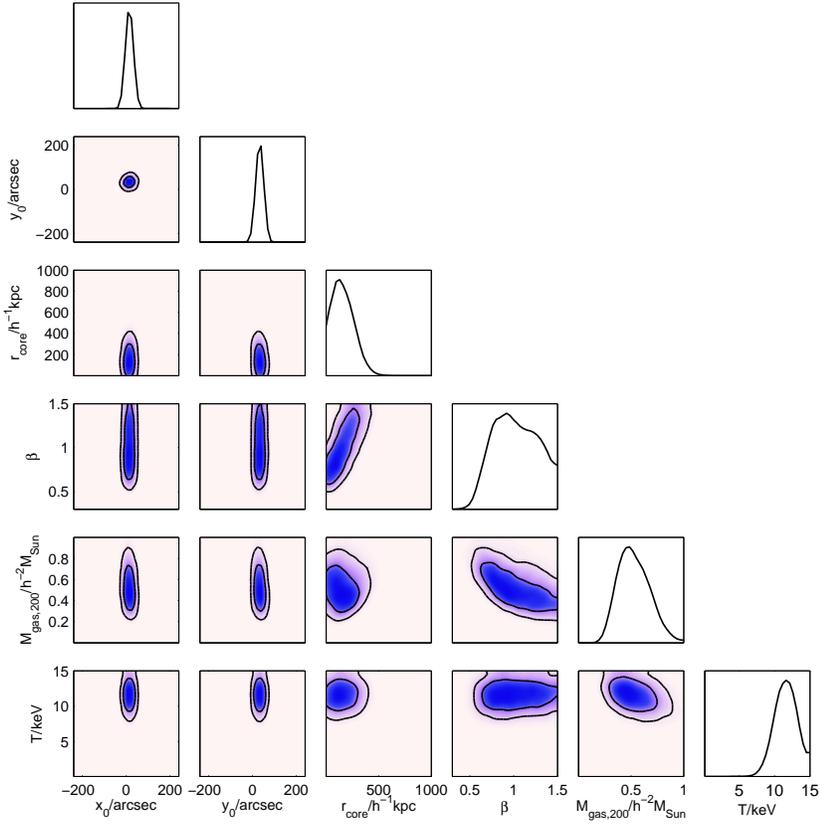}}
\subfigure[For derived parameters, posteriors marginalized into one dimension.]{\includegraphics[trim = 0mm 0mm 0mm 45mm,clip,width=11cm,origin=br,angle=0]{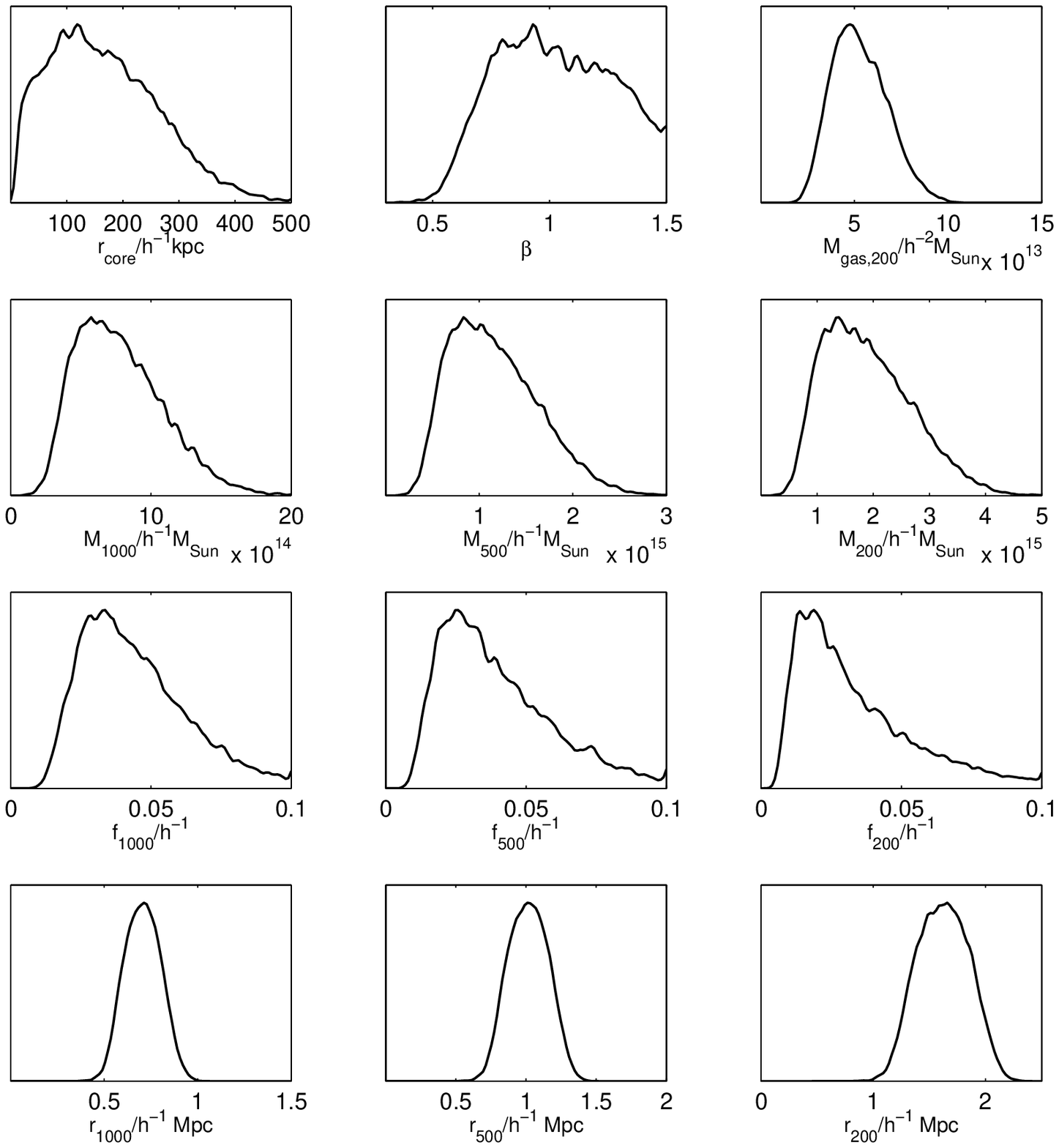}}
\caption{MACSJ0717+37 posterior probability distribution.\label{fig:post:0717}}
\end{figure*}

\begin{figure*}
\centering
\subfigure[For fitted parameters, posteriors marginalized into two dimensions, and into one dimension along the diagonal.]{\includegraphics[width=11cm,origin=br,angle=0]{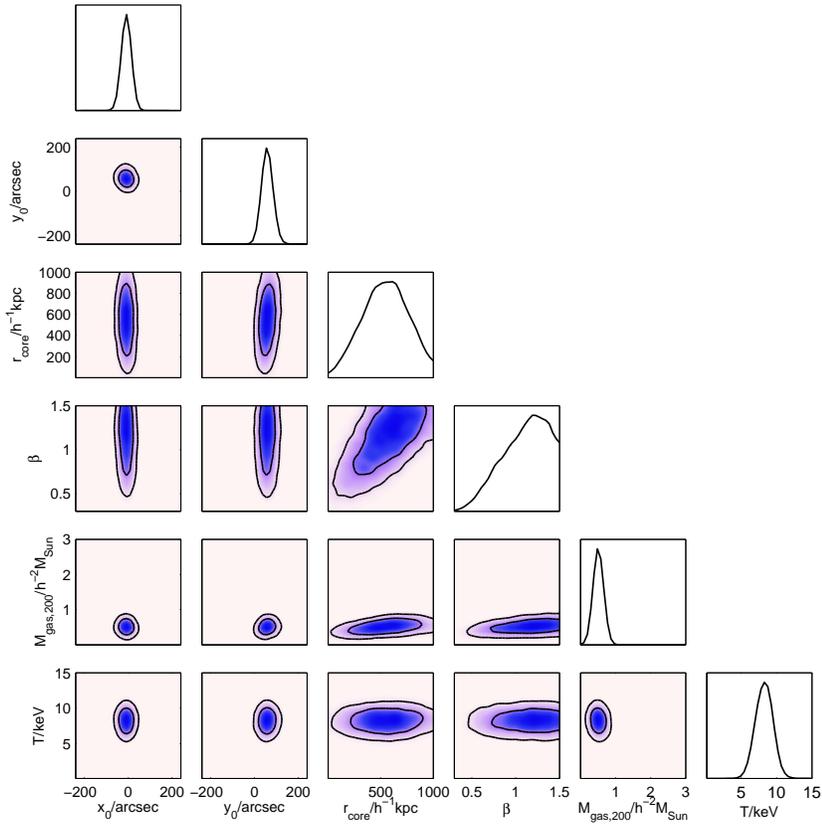}}
\subfigure[For derived parameters, posteriors marginalized into one dimension.]{\includegraphics[trim = 0mm 0mm 0mm 45mm,clip,width=11cm,origin=br,angle=0]{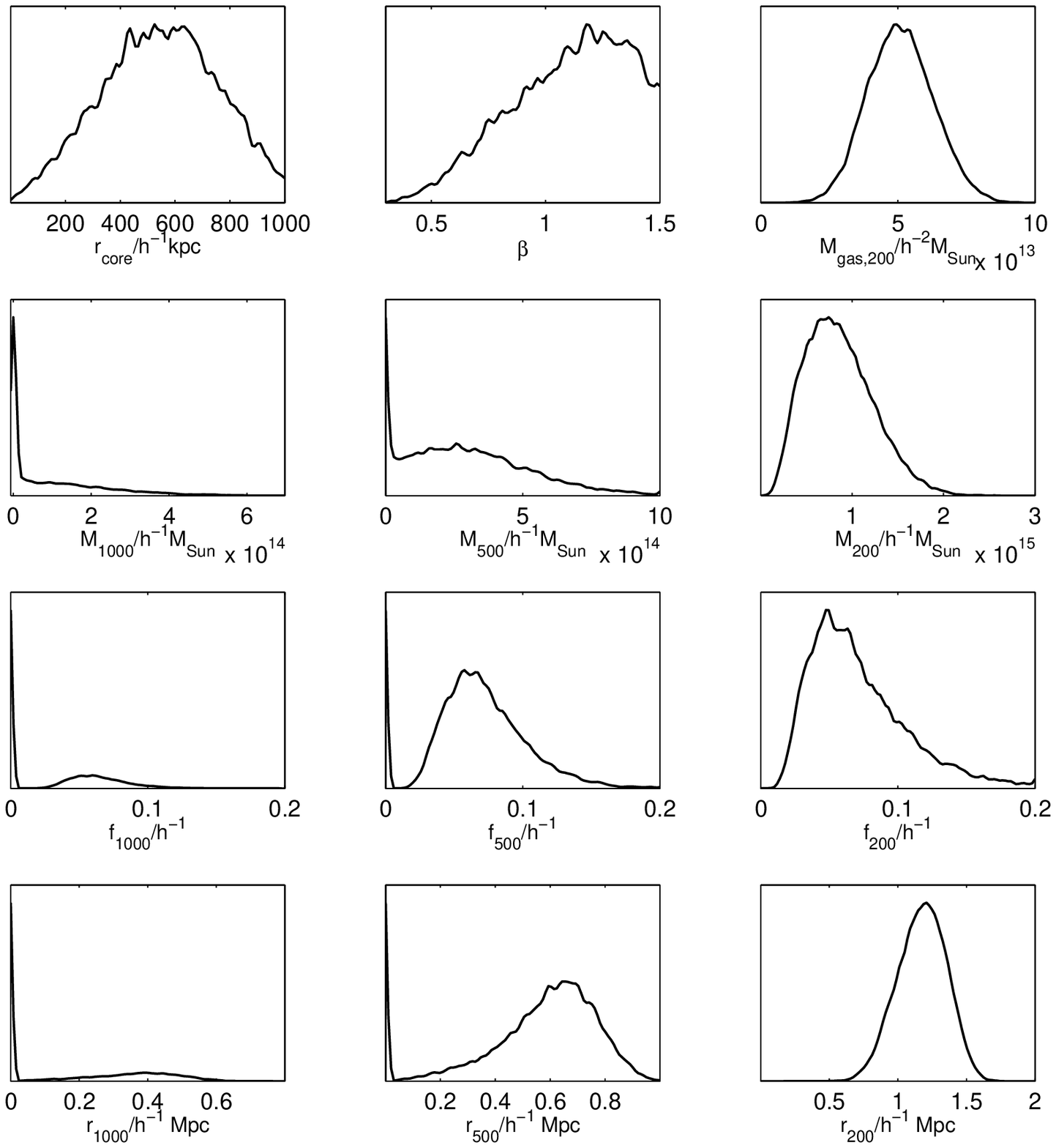}}
\caption{MACSJ0744+39 posterior probability distribution.\label{fig:post:0744}}
\end{figure*}

The full posterior probability distributions for the seven clusters
are shown in Figures \ref{fig:post:a773}--\ref{fig:post:0744}. In each
figure, the upper panel shows the posterior distributions for the
fitted parameters, marginalized into two dimensions, and into one
dimension along the diagonal; the lower panel shows the
one-dimensional marginalized posterior distributions for parameters
derived from those that were fitted. In Table \ref{table:map-params}
we give mean \textit{a posteriori} parameter estimates for the
clusters, but we caution against their use independently of the
posterior probability distributions.

There are two technical points of which to be aware. First, some of
the distributions have rough sections. This roughness is just the
noise due to the finite numbers of samples. We have used narrow
binning of parameter values to avoid misleading effects of averaging
especially at distribution edges, with the consequence of high noise
per bin. Second, there is a possibility that, for some combination of
cluster parameters, nowhere in the cluster does the density reach $a
\times \rho_{\mathrm{crit}}$, resulting in no physical solution for
$r_a$. We set $r_a$ to zero in such cases. Out of the seven clusters
analysed in this paper, this affected only MACSJ0744+39, resulting in
a sharp peak in the posterior probability of $r_{1000}/h^{-1}$Mpc and
$r_{500}/h^{-1}$Mpc close to zero radius. Consequently the posterior
probability also peaks close to zero for derived parameters
$f_{1000}/h^{-1}$, $f_{500}/h^{-1}$, $M_{1000}/h^{-1} M_{\odot}$ and
$M_{500}/h^{-1} M_{\odot}$, for this cluster. A different SA
configuration or more integration would help for MACSJ0744+39, but at
mean overdensity 200 there is no issue.

To set these results in context, we give examples from the literature
of other estimates of some of these quantities that we can find for
these clusters.

For A611, \cite{schmidt07} using \textit{Chandra} find a total virial
mass of $6.2^{+3.8}_{-1.8} \times 10^{14} M_{\odot}$. From gravitational
lensing, \cite{romano10} find $r_{200}$ is some 1.5~Mpc and total mass
is some 4--7$\times 10^{14} M_{\odot}$.

For A773, \cite{zhang08} find from \textit{XMM-Newton} that $r_{500}$
is 1.3~Mpc, $M_{500}$ is $8.3 \pm 2.5 \times 10^{14} M_{\odot}$ and
$f_{g, 500}$ is $0.13 \pm 0.07$, while \cite{barrena07} estimate a
virial mass of $1.2$--$2.7 \times 10^{15} M_{\odot}$ from
\textit{Chandra} and optical-spectral velocities.

For A1914, \cite{zhang08} find from \textit{XMM-Newton} that $r_{500}$
is 1.7~Mpc, $M_{500}$ is $16.8 \pm 4.9 \times 10^{14} M_{\odot}$ and
$f_{g, 500}$ is $0.07 \pm 0.04$. \cite{mrocz09} fit jointly to
\textit{Chandra} and SZA data and find $r_{200}$ is 1.3~Mpc, $M_{500}$
is $6.6$--$8.1 \times 10^{14} M_{\odot}$, and $f_{g, 500}$ is
$0.14$--$0.16$, the exact values depending on assumptions, with random
errors in addition. \cite{zhang10} find from \textit{XMM-Newton} that
$M_{1000}$ is $4.36 \pm 1.22 \times 10^{14} M_{\odot}$ and $M_{500}$
is $7.69 \pm 2.24 \times 10^{14} M_{\odot}$, while from weak lensing
they find that $M_{1000}$ is $3.35^{+0.50}_{-0.47} \times 10^{14}
M_{\odot}$ and $M_{500}$ is $4.46^{+0.75}_{-0.69} \times 10^{14}
M_{\odot}$.

For A2218, \cite{zhang08} find from \textit{XMM-Newton} that
$r_{500}$ is 1.1~Mpc, $M_{500}$ is $4.2 \pm 1.3 \times 10^{14}
M_{\odot}$ and $f_{g, 500}$ is $0.15 \pm 0.09$.

For MACSJ0744+39, \cite{ettori09b} find from \textit{Chandra} that
$r_{200}$ is $1566 \pm 56$~kpc, and also from \textit{Chandra}
\cite{schmidt07} find a virial mass of $7.4^{+4.4}_{-2.1}\times
10^{14}M_{\odot}$.

Returning to our results, three points that are immediately apparent
are that: the gas fractions are low and get lower as $r$ increases; as
well as the usual $\beta$--$r_{\mathrm{core}}$ degeneracy
\citep{grego01,grainge02b,saunders03}, there is a tendency to high
$\beta$; and the results go out to larger radius than typically
obtained from X-ray or SZ cluster analyses. We next consider these
points in more detail.

\subsection{Masses and gas fractions}

Rather than rising towards a canonical large-scale gas fraction of,
say, 0.15 as one goes to large $r$ (see e.g.~\citealt{mccarthy07},
\citealt{komatsu09}, \citealt{ettori09b}), our $f_g$ values are low
and get smaller as $r$ increases. We suspect that our assumption of
isothermality may be the cause. If, away from the central region,
$T_{\mathrm{e}}(r)$ keeps falling as $r$ increases, then of course our
isothermal assumption is invalid. The consequences of this for
estimating $M$ and $f_g$ are however somewhat worse than we initially
expected, for the following reason. In the literature, it is assumed
that the value for $M_{r}$ based on hydrostatic equilibrium (equation
(\ref{eqn:hydro32}) in this work) implies $M_{r} \propto T_{\mathrm{e}}$. But
one has to use equation (\ref{eqn:hydro32}) in terms of radius $r_{a}$
internal to which there is a specific mean overdensity $a$. At a
particular $r_{a}$, one can equate $M_{r}$ from equation
(\ref{eqn:hydro32}) with the expression for $M_{r}$ from integrating
over spherical shells, finding that $r_{a} \propto T^{1/2}$ and in
fact $M_{r} \propto T^{3/2}$ (please note our stated convention at the
end of section \ref{sec:intro}). Since $M_{\mathrm{gas}, r} \propto
T^{-1}$ (given the SZ measurement), $f_{\mathrm{gas}, r}$ is
proportional to $T^{-5/2}$ rather than the $T^{-2}$ in the
literature. It is not possible here to make an approximate
quantitative estimate of the effects of the isothermal assumption
because of its separate effects on $r_c$, on $\beta$, and on total and
gas masses as functions of $r$. Nevertheless, if temperatures are less
than we have assumed, our total mass estimates are biased high, our
gas fraction estimates are biased low, and our $r_a$ estimates are
somewhat biased high.

\subsection{Reaching high radius}

\cite{lacey93} give an expression for how the classical virial radius
($r_{178}$ at $z = 0$) changes with $z$ in an ${\Omega}-{\Lambda}$
universe: for our lowest and highest cluster redshifts, the virial
radii are approximately $r_{205}$ and $r_{215}$. The SA's sensitivity
to structures out to diameters of 10\arcmin corresponds to sensitivity
to a physical diameter of 1.7~Mpc at our lowest cluster redshift.
Given that our $r_{200}$ estimate is biased high, our plots at
overdensity 200 thus reach the virial radius in our nearer clusters
with some extrapolation of the SZ signal and with no extrapolation in
the more-distant ones.

\subsection{$\beta$}

Typical low-$r$ $\beta$-values are about 0.7 (see
e.g.~\citealt{jones84,mohr99,ettori04}) and reach about 0.9 by
$r_{1000}$ (see e.g.~\citealt{vikhlinin99,hallman07}). Despite the
$\beta$--$r_{\mathrm{core}}$ degeneracy, when we marginalize over
everything but $\beta$ we find that $\beta$ is much larger. The two
likely reasons for this are that our data go to high $r$ and that our
estimates of $M_{\mathrm{gas}}$ are biased low at high $r$ because the
$T_{\mathrm{e}}$ we use there is too high; at present we cannot assess the
relative contributions of these two factors.

\begin{figure*}
\centering
\subfigure[For fitted parameters, posteriors marginalized into two dimensions, and into one dimension along the diagonal.]{\includegraphics[width=11cm,origin=br,angle=0]{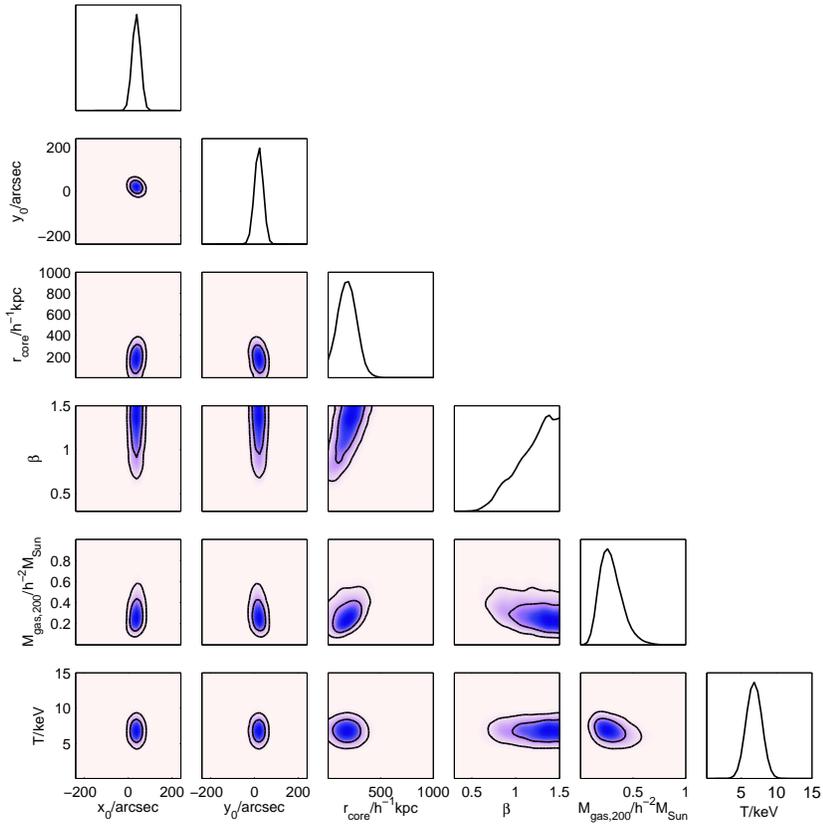}}
\subfigure[]{\includegraphics[trim = 0mm 0mm 0mm 45mm,clip,width=11cm,origin=br,angle=0]{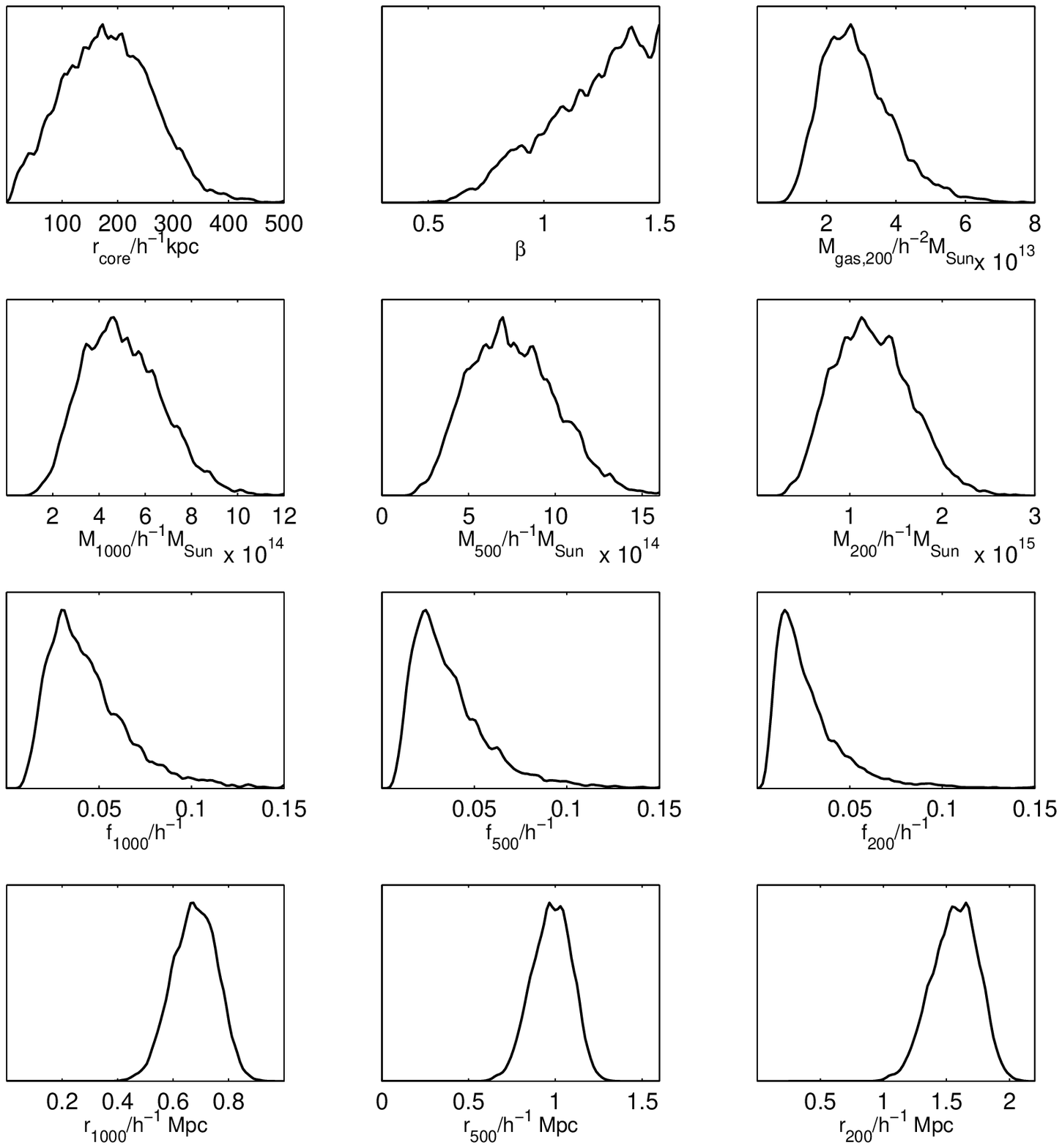}}
\caption{A611 posterior probability distribution.\label{fig:post:a611}}
\end{figure*}

\begin{table*}
\begin{minipage}[h]{14cm}
  \caption{Mean \textit{a posteriori} parameter estimates with 68 per
    cent confidence limits. Note that $e^x$ means $10^{x}$ in this
    table.\label{table:map-params}}
\begin{tabular}{lcccccccc}
\hline
& A611 &   A773 &  A1914 &  A2218 \\
\hline
$r_{\mathrm{core}}/$kpc  &   $2.6e+02^{5.4e+01}_{-6.0e+01}$ &  $3.8e+02^{1.5e+02}_{-1.5e+02}$
&  $1.9e+02^{3.6e+01}_{-3.9e+01}$ &  $3.7e+02^{8.2e+01}_{-5.8e+01}$ \\ 
$\beta$  &   $1.2e+00^{2.9e-01}_{-8.2e-02}$ &  $1.0e+00^{4.9e-01}_{-1.4e-01}$ &
$1.1e+00^{4.3e-01}_{-1.2e-01}$ &  $1.1e+00^{4.1e-01}_{-1.3e-01}$ \\ 
$M_{\mathrm{gas},200}/M_{\odot}$  &   $6.0e+13^{7.1e+12}_{-1.2e+13}$ &
$1.2e+14^{1.6e+13}_{-2.7e+13}$ &  $5.5e+13^{5.3e+12}_{-1.3e+13}$ &
$1.1e+14^{1.3e+13}_{-2.2e+13}$ \\ 
$M_{1000}/M_{\odot}$  &   $7.2e+14^{1.0e+14}_{-1.3e+14}$ &
$6.9e+14^{1.3e+14}_{-1.8e+14}$ &  $1.2e+15^{1.7e+14}_{-2.3e+14}$ &
$7.6e+14^{1.4e+14}_{-1.8e+14}$ \\ 
$M_{500}/M_{\odot}$  &   $1.1e+15^{1.6e+14}_{-2.0e+14}$ &
$1.1e+15^{2.1e+14}_{-2.8e+14}$ &  $1.7e+15^{2.5e+14}_{-3.5e+14}$ &
$1.2e+15^{2.4e+14}_{-2.8e+14}$ \\ 
$M_{200}/M_{\odot}$  &   $1.8e+15^{2.7e+14}_{-3.2e+14}$ &
$1.9e+15^{3.7e+14}_{-4.8e+14}$ &  $2.7e+15^{4.0e+14}_{-5.6e+14}$ &
$2.0e+15^{4.2e+14}_{-4.8e+14}$ \\ 
$r_{1000}/$Mpc  &   $6.8e-01^{4.1e-02}_{-3.7e-02}$ &  $5.3e-01^{4.7e-02}_{-3.9e-02}$ &
$8.3e-01^{5.1e-02}_{-4.9e-02}$ &  $7.1e-01^{5.7e-02}_{-4.7e-02}$ \\ 
$r_{500}/$Mpc  &   $9.8e-01^{5.9e-02}_{-5.2e-02}$ &  $6.8e-01^{5.7e-02}_{-4.9e-02}$ &
$1.2e+00^{7.4e-02}_{-7.1e-02}$ &  $1.0e+00^{8.9e-02}_{-6.7e-02}$ \\ 
$r_{200}/$Mpc  &   $1.6e+00^{9.6e-02}_{-8.1e-02}$ &  $1.0e+00^{8.6e-02}_{-7.1e-02}$ &
$1.9e+00^{1.2e-01}_{-1.1e-01}$ &  $1.7e+00^{1.5e-01}_{-1.1e-01}$ \\ 
$f_{1000}$  &   $6.4e-02^{7.5e-03}_{-1.9e-02}$ &  $9.5e-02^{1.2e-02}_{-2.6e-02}$ &
$3.8e-02^{4.1e-03}_{-1.3e-02}$ &  $9.5e-02^{9.5e-03}_{-3.0e-02}$ \\ 
$f_{500}$  &   $5.3e-02^{5.7e-03}_{-1.8e-02}$ &  $9.5e-02^{1.2e-02}_{-3.1e-02}$ &
$3.3e-02^{2.8e-03}_{-1.3e-02}$ &  $8.9e-02^{5.2e-03}_{-3.6e-02}$ \\ 
$f_{200}$  &   $4.1e-02^{2.8e-03}_{-1.7e-02}$ &  $9.5e-02^{8.5e-03}_{-4.0e-02}$ &
$2.7e-02^{1.2e-03}_{-1.3e-02}$ &  $8.3e-02^{4.3e-03}_{-4.4e-02}$ \\
\hline
\end{tabular}
\end{minipage}
\end{table*}

\begin{table*}
\begin{minipage}[h]{14cm}
\begin{tabular}{lcccccccc}
\hline
& MACSJ0308+26 &  MACSJ0717+37 & MACSJ0744+39 \\
\hline
$r_{\mathrm{core}}/$kpc  &   $8.6e+02^{8.3e+01}_{-6.7e+01}$ &  $2.4e+02^{5.7e+01}_{-8.3e+01}$
&  $7.8e+02^{1.5e+02}_{-1.5e+02}$ \\ 
$\beta$  &   $1.2e+00^{2.6e-01}_{-6.2e-02}$ &  $1.0e+00^{4.8e-01}_{-1.4e-01}$ &
$1.1e+00^{4.0e-01}_{-1.1e-01}$ \\ 
$M_{\mathrm{gas},200}/M_{\odot}$  &   $1.8e+14^{1.6e+13}_{-2.0e+13}$ &
$1.1e+14^{1.3e+13}_{-1.7e+13}$ &  $1.0e+14^{1.1e+13}_{-1.2e+13}$ \\ 
$M_{1000}/M_{\odot}$  &   $7.0e+14^{1.4e+14}_{-7.0e+14}$ &
$1.1e+15^{1.7e+14}_{-2.6e+14}$ &  $1.3e+14^{1.8e+14}_{-1.3e+14}$ \\ 
$M_{500}/M_{\odot}$  &   $1.7e+15^{2.4e+14}_{-3.2e+14}$ &
$1.7e+15^{2.7e+14}_{-4.1e+14}$ &  $4.4e+14^{1.2e+14}_{-4.4e+14}$ \\ 
$M_{200}/M_{\odot}$  &   $3.4e+15^{4.9e+14}_{-5.6e+14}$ &
$2.7e+15^{4.5e+14}_{-6.8e+14}$ &  $1.2e+15^{2.1e+14}_{-3.0e+14}$ \\ 
$r_{1000}/$Mpc  &   $6.4e-01^{6.4e-02}_{-4.4e-02}$ &  $7.1e-01^{4.8e-02}_{-4.9e-02}$ &
$2.1e-01^{1.5e-01}_{-2.1e-01}$ \\ 
$r_{500}/$Mpc  &   $1.1e+00^{6.8e-02}_{-6.0e-02}$ &  $1.0e+00^{7.2e-02}_{-7.3e-02}$ &
$5.7e-01^{1.1e-01}_{-5.7e-01}$ \\ 
$r_{200}/$Mpc  &   $1.9e+00^{1.1e-01}_{-9.0e-02}$ &  $1.6e+00^{1.2e-01}_{-1.2e-01}$ &
$1.2e+00^{9.2e-02}_{-8.1e-02}$ \\ 
$f_{1000}$  &   $8.4e-02^{9.2e-03}_{-1.9e-02}$ &  $6.5e-02^{8.4e-03}_{-1.8e-02}$ &
$5.5e-02^{3.0e-02}_{-5.5e-02}$ \\ 
$f_{500}$  &   $7.5e-02^{7.8e-03}_{-1.9e-02}$ &  $5.9e-02^{7.9e-03}_{-2.0e-02}$ &
$1.0e-01^{1.6e-02}_{-1.0e-01}$ \\ 
$f_{200}$  &   $6.1e-02^{5.8e-03}_{-1.9e-02}$ &  $5.2e-02^{5.8e-03}_{-2.2e-02}$ &
$1.0e-01^{1.3e-02}_{-3.2e-02}$ \\
\hline
\end{tabular}
\end{minipage}
\end{table*}

\section{Conclusions}
\label{conclusions}

\begin{enumerate}

\item{Untapered, naturally-weighted AMI Small Array maps at
    13.9--18.2~GHz, with no source subtraction, show clear SZ effects
    in five of the seven clusters.}

\item{Using source-subtraction observations that are largely from the
    Ryle Telescope (and thus at $15$~GHz but typically two years
    before the SA observations), and assuming a spherical
    $\beta$-model, hydrostatic equilibrium, and isothermality with an
    X-ray measured temperature, our Bayesian analysis reveals SZ
    signals in all seven clusters. In six of these, the Bayesian
    evidence for an SZ detection, in addition to sources plus CMB
    primary anisotropies plus thermal noise, is huge; in the one of
    them with much the worst thermal noise, there is a 1 in 3000
    chance that the SZ is spurious. We emphasize that, to allow for
    variability, we set the prior on each source's flux density as its
    high-resolution value with a Gaussian 1-$\sigma$ width of (except
    in one case) $\pm 30$~per~cent.}

\item{The Bayesian evidence proves very useful in understanding source
    environments. For example, a high-resolution map showed a feature
    that, by eye, was classed as a tentative radio-source
    detection. Running the Bayesian analysis twice, with and without
    that tentative source, showed that the evidence for it is in fact
    so low that it should not be included.}

\item{We note that our sensitivity to structures out to 10\arcmin,
    corresponding to a 1.7-Mpc diameter for our lowest-redshift
    cluster, means that our parameter estimates out to the classical
    virial radii of the nearer clusters involve some extrapolation,
    but no extrapolation is needed for the more-distant ones.}

\item{Our probability distributions of masses and radii internal to
    which the average overdensities are 1000, 500 and 200 are usefully
    constrained and change sensibly over this range. \emph{However},
    our gas fractions are evidently low compared with values in the
    literature; further, they decrease with increasing radius, which
    is also unexpected. The problem seems consistent with the notion
    that temperature $T_{\mathrm{e}}$ decreases as radius $r$
    increases whereas we are assuming issothermality (using
    temperatures measured from low-radii data); the problem is made
    somewhat worse because, as we have shown, gas fraction goes as
    $T_{\mathrm{e}}^{-2.5}$ assuming isothermality and hydrostatic
    equilibrium rather than as $T_{\mathrm{e}}^{-2}$ as seems to have
    been assumed in the literature. If $T_{\mathrm{e}}$ does indeed
    fall as $r$ increases, our gas masses are biased low and our total
    masses (and to a lesser extent our measurements of $r_a$) are
    biased high. Temperature profiles must be measured or some other
    means found to deal with this problem if we are to infer masses
    out towards the virial radius. Indeed, along with other
    density-profile models, this will be investigated in future work.}

\end{enumerate}

\section*{Acknowledgments}
\label{ack}

We thank PPARC/STFC for support for AMI and its operation. We thank
PPARC for support for the RT and its operation. We warmly acknowledge
the staff of Lord's Bridge and of the Cavendish Laboratory for their
work on AMI and the RT. MLD, TMOF, MO, CRG and TWS acknowledge
PPARC/STFC studentships. The analysis work was conducted in
cooperation with SGI/Intel using the Altix 3700 supercomputer at
DAMTP, University of Cambridge supported by HEFCE and STFC, and we are
grateful to Andrey Kaliazin for computing assistance.
 
\bibliographystyle{mn2e}
\bibliography{ami-one}\label{lastpage}


\end{document}